\documentclass[12pt]{iopart}
\newcommand{\gguide}{{\it Preparing graphics for IOP journals}}
\usepackage{iopams}
\usepackage[dvips]{graphicx}
\usepackage{amssymb}
\usepackage{psfrag}
\usepackage{color}
\usepackage{subfigure}
\usepackage{sidecap}
\def\RR{ \mathbb R}

\def\bl{ \mathbf \lambda}

\def\bt{\boldsymbol{\theta}}
\def\bl{\boldsymbol{\Lambda}}

\usepackage{epsf}
\usepackage{harvard}
\newcommand{\eee}{\end{equation}}
\newcommand{\bee}{\begin{equation}}
\newcommand{\ec}{\end{center}}
\newcommand{\bc}{\begin{center}}
\newcommand{\eea}{\end{eqnarray}}
\newcommand{\bea}{\begin{eqnarray}}
\newcommand{\bd}{\begin{description}}
\newcommand{\ed}{\end{description}}
\newcommand{\bbi}{\begin{itemize}}
\newcommand{\eei}{\end{itemize}}
\newcommand{\pa}{\partial}
\newcommand{\refeq}[1]{Equation (\ref{#1})}

\begin{document}

\title[P.S. Koutsourelakis]{
A novel Bayesian strategy for the identification of spatially-varying material properties  and model validation: an application to static elastography}

\author{P.S. Koutsourelakis}

\address{Center for Applied Mathematics, Cornell University}
\ead{pk285@cornell.edu}
\begin{abstract}
The present paper proposes a novel Bayesian, computational strategy  in the context of model-based inverse problems in   elastostatics. On one hand we attempt to provide probabilistic estimates of the material properties and their spatial variability that account for the various sources of uncertainty. On the other hand we attempt to address the question of model fidelity in relation to the experimental reality and  particularly in the context of the material constitutive law adopted. This is especially important in biomedical settings when the inferred material properties will be used to make decisions/diagnoses. We propose an expanded parametrization that enables the quantification of model discrepancies in addition to the constitutive parameters. We propose scalable computational strategies for carrying out inference and learning tasks and demonstrate their effectiveness in numerical examples with noiseless and noisy synthetic data.

\end{abstract}

\maketitle

\section{Introduction}
\label{sec:intro}

%

The extensive use of large-scale computational models  poses several challenges in parameter identification in the context of system identification or performing predictive simulations.
Medical imaging represents such an application which has attracted significant interest in recent years as the correct identification of material properties can reveal  various pathologies \cite{ISI:000263259100006,liver2006} as well as quantitatively assess the progress of various treatments.



Ultrasound elasticity imaging (elastography)  has gained prominense  in the context of performing medical diagnosis due its accuracy and low cost. 
It is based on ultrasound tracking of pre- and post-compression images to obtain a map of position changes from which deformations can be inferred.  
The pioneering work of   Ophir and coworkers \cite{Ophir:1991}
 followed by several clinical studies \cite{Garra:1997,Bamber:2002,Hall:2003,Giuseppetti:2005,Itoh:2006,Thomas:2006,Regner:2006,Burnside:2007,Zhi:2007,par11im} have demonstrated that the resulting strain images
typically improve the diagnostic accuracy over ultrasound alone.



Broadly speaking, there are two approaches that are utilized for calculating the
constitutive parameters. In the direct approach, the equations
of equilibrium  are interpreted as  equations for the material parameters of interest, where the inferred strains and their derivatives appear as coefficients \cite{ISI:000275699600006,ISI:000237030900021,alb09adj}. While such an approach provides a computationally efficient strategy that does not  require solution over the whole domain nor knowledge of the boundary conditions,  it has certain drawbacks.
More importantly perhaps,  it does not use the raw data (i.e. noisy displacements) but transformed versions i.e. strain fields which arise by applying sometimes ad hoc filtering and  smoothing operators. While these might be plausible, in general alter the informational content of the data and make difficult the quantification of the effect of observation noise.
 This is amplified when strain derivatives are computed, although  not all such approaches require them e.g.  \cite{ISI:000267195700003}. Furthermore, the smoothing employed can smear regions with sharply varying properties and hinder proper identification.
Finally, it is non-trivial
 to  determine appropriate boundary
conditions in terms of the material parameters of interest.

The alternative to direct methods, i.e. indirect, or iterative, as they are most commonly referred to, admit an inverse problem formulation where the discrepancy (in various norms, \cite{ISI:000240849100006,ISI:000255220100005}) between observed and model-predicted displacements is minimized with respect to the material fields of interest \cite{ISI:000237981300009,ISI:000223500200013,ISI:000238422400023,lakisorig,ISI:000232236800008,ISI:000228126000006,ISI:000262358700008,ISI:000257837900010,ISI:000275756200016,ISI:000280774700004,bon05inv}.
 While these approaches utilize directly the raw data, they generally imply an increased computational cost as the forward problem and potentially derivatives have to be solved/computed several times.
This effort is amplified when stochastic/statistical formulations are employed as those arising from the Bayesian paradigm, whose cost is comparable to that of a deterministic global optimization technique \cite{kai05com}.

Bayesian techniques are advocated in this paper  due their ability to quantify the effect of various sources of uncertainty to the hypotheses tested or the inferences made.
 One source of uncertainty is obviously the noise in the data which constitutes probabilistic estimates more rational. This is particularly important when multiple hypotheses are consistent with the data or the level of confidence in the estimates produced needs to be quantified.
Another source of uncertainty which is largely  unaccounted for, is {\em model  uncertainty} \cite{Higdon:2008}. Namely, the parameters, whose values are estimated, are associated with a particular forward model about the behavior of the medium  (in our case a system of PDEs consisting of equilibrium and constitutive equations) but one cannot be certain about the validity of the model employed. In general, there will be deviations between the physical reality where measurements are made, and the idealized mathematical/computational description.
Especially in the context of medical applications, it is crucially important  to account for the model  discrepancy or inadequacy in order to infer the right material properties and make accurate diagnoses. \footnote{"I remember my friend Johnny von Neumann used to say, 'with four parameters I can fit an elephant and with five I can make him wiggle his trunk.'" {\em A meeting with Enrico Fermi}, Nature 427, 297; 2004.}
Non-intrusive Bayesian strategies, i.e. those that basically make use of the forward model as a black-box,   capture  model discrepancy with regression models (e.g. Gaussian processes) which are not easily physically-interpretable and cumbersome or impractical when they depend  on a large number of  input parameters. \cite{Kennedy:2001,Higdon:2008}.
In contrast, our approach is intrusive. This enables us to overcome the aforementioned limitations and allows us to directly infer the stresses/pressure in the context of {\em elastostatics}.

The rest of the paper is organized as follows. Section \ref{sec:meth} is devoted to the presentation of the novel Bayesian framework proposed in the context of elastostatics. Section \ref{sec:inference} discusses  computational aspects related to inference techniques for sampling from the posterior and learning schemes for estimating parameter values. Finally section \ref{sec:examples}  presents numerical results under static plane stress conditions using noiseless and noisy data with particular emphasis on quantifying model discrepancy.

\section{Proposed Methodology}
\label{sec:meth}

  The presentation of the ideas in this paper is centered around solid mechanics, in particular elastostatics, but the framework introduced can be directly transitioned to other  continua. We discuss first the formulation of the probabilistic model proposed and in subsection \ref{sec:inference} the inference and learning tasks associated with this description. 
We adopt a physically-inspired strategy that focuses on quantifying model discrepancies in the context of the {\em constitutive equation}.
From a deterministic point of view it resembles techniques such as constitutive relation error (CRE) or error in the constitutive equation (ECE) that have been developed for a  posteriori error estimation and the solution of overspecified inverse problems \cite{V1984,P1999,bon05inv,A2004a,P2007}. 
We use the term constitutive equations to refer in general to relations between conjugate thermodynamic variables, i.e.  stress and strain in solid mechanics or  velocity and pressure in flow through permeable media or flux and temperature in heat diffusion. 

{\em In the formulations proposed, the constitutive relation    supplements the observables and an augmented state space is used that includes all conjugate variables.}  As it is demonstrated in the sequence, the addition of these  unknown parameters simplifies inference tasks and enables the quantification of model errors.
The motivation for such an approach stems from the fact that inverse problems in the context of continuum models consist  of:
\bbi
\item  a {\em conservation law} that arises from physical principles that are generally well-founded and trusted. In the case of single-phase flow through a porous medium this amounts to the conservation of mass,  in 
solid  mechanics  to  the conservation of linear momentum. In elastostatics in particular 
this is written as:
\bee
\label{eq:solidmech}
\nabla \cdot \mathbf{\tilde{\sigma}}(\mathbf{x}) +\mathbf{b}(\mathbf{x})= \mathbf{0}, \quad \mathbf{x} \in \Omega
\eee
where $\mathbf{\tilde{\sigma}}(\mathbf{x})$ is the stress tensor, $\mathbf{b}$ the body force and $\Omega$ the problem domain.
Discretized versions of the aforementioned PDE are employed which naturally introduce {\em discretization} error. This is generally well-studied in the context of linear problems and several a priori (and a posteriori) error estimates are available. In this work we will ignore the discretization error in \refeq{eq:solidmech} which corresponds to the verification stage and focus on the validation and calibration aspects. 
\item  a {\em constitutive law} that is by-and-large phenomenological and therefore provides the primary source of {\em model uncertainty}. This is represented by the conductivity tensor in heat diffusion,  the permeability tensor in flow through porous media or 
the elasticity tensor $\mathbf{D}$ in solid mechanics:
\bee
\label{eq:constelast}
\boldsymbol{\sigma}(\mathbf{x})= \mathbf{D}(\mathbf{x}) \boldsymbol{\epsilon}(\mathbf{x}), \quad \forall \mathbf{x} \in \Omega 
\eee
where $\boldsymbol{\sigma}(\mathbf{x})$ is the vector of stresses  and  $\boldsymbol{\epsilon}(\mathbf{x})$ the vector of strains.
\item boundary/initial conditions or observables in general (which might include interior displacements). The available data are  contaminated by noise and represent the main source of  {\em observation errors}.
\eei
In the Bayesian setting advocated, the  goal is to evaluate the {\em posterior density} for the material parameters (i.e.  $\mathbf{D}(\mathbf{x})$)  as well as quantitatively assess the validity of the aforementioned constitutive relation (\refeq{eq:constelast}).

The numerical implementation requires discretization of the aforementioned equations.  For economy of notation, we consider the simplest perhaps discretization consiting of  a finite element triangulation $\mathcal{T}$ of the problem domain $\Omega$ using $n_{el}$  constant-strain/stress elements \footnote{For more complex elements/discretizations, the ensuing formulations can be readily applied if instead we consider each integration point in the element}. 
If   $e$ denotes the element number, the parameters in the formulation proposed are:
\bbi
\item the   stress vectors $\boldsymbol{\sigma}_e, ~e=1,\ldots, n_{el}$  ( $3-$ dimensional under plane stress/strain conditions or $6$-dimensional in general three-dimensional problems) which are jointly denoted by $\boldsymbol{\sigma}=\left[\boldsymbol{\sigma}_1,\ldots, \boldsymbol{\sigma}_{n_{el}} \right]^T$.
\item the global displacement vector $\mathbf{u}$. If $\mathbf{u}_e$ denotes the  nodal displacement vector of element $e$ then  we represent by $\mathbf{L}_e$ the Boolean  matrices that  relate local and global displacement vectors i.e. $\mathbf{u}_e=\mathbf{L}_e \mathbf{u}$. We further denote by $\boldsymbol{\epsilon}_e$ the element strain vector which is related to $\mathbf{u}_e$ as $\boldsymbol{\epsilon}_e=\mathbf{B}_e \mathbf{u}_e$ where $\mathbf{B}_e$ is the well-known  strain-displacement matrix.
\item the local constitutive matrices $\mathbf{D}_e$ that relate stress and strains over element $e$, i.e. $\boldsymbol{\sigma}_e= \mathbf{D}_e \boldsymbol{\epsilon}_e$. These are assumed constant over each element but they could be assigned different values at the nodes of the mesh  or integration points of each element.

\eei

We will further assume that {\em noisy} displacement data (at interior or boundary points) are provided and will be denoted by $\mathbf{u}_Q \in \RR^{n_Q}$. It is assumed that the observed nodal displacements are given by $\mathbf{Q}~\mathbf{u}$ where $\mathbf{Q}$ is an appropriate Boolean matrix (if all displacements are observed at all the nodes then $\mathbf{Q}=\mathbf{I}$). Assuming Gaussian noise with variance $\nu^2$, the {\em likelihood} of $\mathbf{u}_Q$ given $\mathbf{u}$ is normal and:
\bee
\label{eq:disp}
p(\mathbf{u}_Q \mid \mathbf{u}) \propto \frac{1}{\nu^{n_Q}} \exp\{-\frac{1}{2\nu^2} (\mathbf{u}_Q- \mathbf{Q}\mathbf{u})^T(\mathbf{u}_Q- \mathbf{Q}\mathbf{u}) \}
\eee
The observation noise variance $\nu^2$ can be known or unknown in which case we propose employing a conjugate $inverse-Gamma$ hyperprior with hyperparameters $(\alpha_{\nu}, \beta_{\nu})$, i.e:
\bee
\label{eq:priornu}
p(\nu^2)\propto (\nu^{-2})^{\alpha_{\nu}-1} e^{-\beta_{\nu}/\nu^2}
\eee
 Naturally more complex models that can capture perhaps  the  spatial dependence of $\nu$ can be employed.
In general, non-essential boundary conditions might be available as well, i.e. tractions might be prescribed at part of the boundary $\pa \Omega_N \subset \pa \Omega$ i.e.:
\bee
\mathbf{n} \cdot \tilde{\sigma}(\mathbf{x}) \mid_{\Omega_N}=\boldsymbol{\tau}(\mathbf{x}), \quad \mathbf{x} \in  \pa \Omega
\eee
Noise in these observations could also  be added but we omit this to simplify notation.

In the proposed framework, apart from the aforementionned observations,  the {\em  data}  or likelihood consist also of model-related  equations, i.e. the conservation law (\refeq{eq:solidmech}) which in the case of standard Bubnov-Galerkin finite element schemes is enforced weakly as: 
\bee
\label{eq:deq1}
\int_{\Omega}  \boldsymbol{\epsilon}(\mathbf{w}) \cdot \boldsymbol{\sigma} d\mathbf{x}=\int_{\Omega}\mathbf{w} \cdot \mathbf{b}~d\Omega+\int_{\pa \Omega_N} \mathbf{w} \cdot \boldsymbol{\tau} ~d\Gamma
\eee
where  $\boldsymbol{\epsilon}(\mathbf{w})$ denote the strains associated with the the weighting functions $\mathbf{w} \in H^1_0(\Omega)$. 
It is noted that other discretization schemes such as finite volume or discontinuous Galerkin  can also be used to enforce the conservation law. 
with small alterations. In the triangulation $\mathcal{T}$ adopted for discretizing the solution and the weighting functions $\mathbf{w}$ this reduces to:
\bee
\label{eq:deq}
\mathbf{\hat{B}}^T \boldsymbol{\sigma}=\mathbf{f}
\eee
where $\mathbf{f}$ is the force vector and:
\bee
\mathbf{\hat{B}}^T=\sum_{e=1}^{n_{el}} (\mathbf{L}_e)^T \int_{\Omega_e} (\mathbf{B}_e)^T~d\mathbf{x}=\sum_{e=1}^{n_{el}}  V_e (\mathbf{L}_e)^T (\mathbf{B}_e)^T
\eee
where  $V_e$ is the volume of element $e$.

The second model equation relates to the constitutive law which we propose enforcing  for every element probabilistically. If 
the true constitutive law (which is unknown) is different from the one prescribed in \refeq{eq:constelast}, then there will be a discrepancy/error  $\mathbf{c}_e$ between 
 the actual stresses $\boldsymbol{\sigma}_e$ and the model-predicted stresses $\mathbf{D}_e \boldsymbol{\epsilon}_e= \mathbf{D}_e \mathbf{B}_e \mathbf{u}_e$:
\bee
\label{eq:ece}
 \mathbf{c}_e=\boldsymbol{\sigma}_e-\mathbf{D}_e \mathbf{B}_e \mathbf{u}_e
\eee
Since $\mathbf{c}_e$ is unknown and in accordance  with the Bayesian formulation advocated, we propose a hierarchical prior model where:
\bee
\label{eq:pece}
\begin{array}{l}
\mathbf{c}_e \mid \boldsymbol{\sigma}_e, \mathbf{u}_e,\mathbf{\Sigma}_e \sim  \mathcal{N}(\boldsymbol{\sigma}_e-\mathbf{D}_e \mathbf{B}_e \mathbf{u}_e, \mathbf{\Sigma}_e) \\
or \\
p({c}_e \mid \boldsymbol{\sigma}_e, \mathbf{u}_e,\mathbf{\Sigma}_e) \propto \frac{1}{\mid \mathbf{\Sigma}_e \mid ^{1/2} } exp\{ -\frac{1}{2} (\boldsymbol{\sigma}_e-\mathbf{D}_e \mathbf{B}_e \mathbf{u}_e)^T\mathbf{\Sigma}_e^{-1}  (\boldsymbol{\sigma}_e-\mathbf{D}_e \mathbf{B}_e \mathbf{u}_e) \}
\end{array}
\eee

In this work we consider a special form of the covariances $\mathbf{\Sigma}_e =\lambda_e^2 \mathbf{I}$.
The hyperparameters $\lambda_e^2$  express the variability of the constitutive error  and their magnitude quantifies the {\em model discrepancy}  over each element $e$. The inferred values $\lambda_e^2$ will reveal elements where the model error is high and  refinement/improvement is needed. Note for example that if the elastic properties vary within an element $e$, the corresponding $\lambda_e^2$ will be non-zero even if no noise exists in the data. When different discretization schemes are used which might  employ higher-order shape functions,    distinct $\lambda_e^2$ for each integration point can be introduced. 
The normal prior for $\mathbf{c}_e$ (\refeq{eq:pece}) is not the only option and was selected here for computational convenience due to its conjugacy with the other distributions as it will be seen in the sequel. It would certainly be worth-while to investigate alternative prior models.

Since the hyperparameters $\lambda_e^2$ are unknown, prior models can be employed as well. In this study we make of a Gaussian Markov Random Field (GMRF, \cite{bes91bay,bes93spa}) prior which accounts for the fact that the magnitude of the model errors are expected to be spatially correlated. In particular, and since $\lambda_e^2 \ge 0$ we define the prior implicitly through the vector $\mathbf{Z}=\{z_e\}_{e=1}^{n_{el}}$ where $z_e =\log \lambda_e^2$:
\bee
\label{eq:priorl}
p(\mathbf{\Lambda}) \propto  \exp\{ -\frac{1}{2} \mathbf{Z}^T \mathbf{W}\mathbf{Z} \}
\eee
The precision matrix is given by $\mathbf{W}=\frac{1}{\sigma_z^2}\mathbf{H}$ 
where $\sigma_z^2$ is a scale parameter 
 and  $\mathbf{H}=[H_{e_1,e_2}]$:
\bee
\label{eq:h}
H_{e_1,e_2}=\left\{ \begin{array}{ll}
    \sum_{e_2=1}^{n_{el}} h_{e_1,e_2} & \textrm{if } e_1=e_2 \\
    -h_{e_1,e_2} & \textrm{otrherwise}             
               \end{array} \right.
\eee
where  $h_{e_1,e_2}>0$ is a mesure of proximity between elements $e_1$ and $e_2$. In this work this was defined with respect to the distance $d_{e_1,e_2}$ between the element centroids  as $ h_{e_1,e_2}=e^{-d_{e_1,e_2}/d_0}$ where $d_0$ is a correlation-length parameter. 
The aforementioned model represents an  intrinsic
autoregressive prior \cite{kun87int,bes95con}, which is an improper
distribution (since $\mathbf{W}$ is semi-positive definite) that has been extensively used in spatial statistics. In particular, since $\sum_{e_2} W_{e_1,e_2}=0~ \forall e_1$, it can be easily established that $p(\mathbf{\Lambda})$ penalizes the ``jumps'' in $\mathbf{Z}$ at neighboring elements, i.e.:
\bee
p(\mathbf{\Lambda})\propto \exp\{ \sum_{e_1<e_2} W_{e_1,e_2} (z_{e_1}-z_{e_2})^2 \}
\eee
It is noted finally that values for the parameters $(d_0, \sigma_z^2)$ are provided in the numerical results section.



The combination of Equations (\ref{eq:disp}), (\ref{eq:priornu}),  (\ref{eq:deq}), (\ref{eq:pece}) and (\ref{eq:priorl}) leads to the {\em posterior density} on the model parameters $\mathbf{\Theta} = \left(\nu^2, \boldsymbol{\sigma}, \{\mathbf{D}_e\}_{e=1}^{n_{el}}, \mathbf{u}, \boldsymbol{\Lambda}=\{\lambda_e^2\}_{e=1}^{n_{el}}\right)$. In addition to the obervations $\mathbf{u}_Q$, the posterior on $\mathbf{\Theta}$ is {\em explicitly} conditioned on the model equations i.e. the discretized equation of equilibrium and the constitutive law \footnote{this conditioning is denoted by $\mathcal{M}$ in \refeq{eq:post}}:
\bee
\label{eq:post}
\begin{array}{ll}
\pi(\mathbf{\Theta})=p(\mathbf{\Theta} \mid \mathbf{u}_Q, \mathcal{M})  = & p(\mathbf{u}_Q \mid \mathbf{u}, \nu^2) p(\nu^2) \\
& 1_{\{\mathbf{\hat{B}}^T \boldsymbol{\sigma}=\mathbf{f}\}}(\boldsymbol{\Theta}) \\ & \prod_{e=1}^{n_{el}} p(\mathbf{c}_e \mid \boldsymbol{\sigma}_e, \mathbf{u}_e, \lambda_e^2) ~p(\boldsymbol{\Lambda}) \\
& p(\mathbf{u})
\end{array}
\eee
The indicator function $1_{\{\mathbf{\hat{B}}^T \boldsymbol{\sigma}=\mathbf{f}\}}(\boldsymbol{\Theta})$ implies that the support of the distribution includes only stress vectors that satisfy the (discretized) equilibrium equations in \refeq{eq:deq}.

A prior model could also be adopted with repsect to the constitutive parameters $\mathbf{D}_e$. Such priors apart from improving the regularity of posterior are also physcially plausible as one would expect the constitutive properties at neighboring locations to be correlated. 
Naturally several such models have been proposed in the literature \cite{kai05com}. 
In this work however this was found unnecessary as the formulation  proposed  provides a natural correlation between $\mathbf{D}_e$ through the dispacements $\mathbf{u}$ and stresses $\boldsymbol{\sigma}$ which are themselves spatially correlated due to the equilibrium and constitutive equations. This is evident in the conditional posteriors presented in the sequence.
In contrast a  prior model was adopted for the displacement vector, denoted by $p(\mathbf{u})$ in \refeq{eq:post}. This can be useful when the  observed displacements are sparse or restricted to a portion of the problem domain but its primary utility in the examples contained in section \ref{sec:examples} was found to be the regularization of the displacement field in the presence of noise. In particular we adopted an intrinsic autoregressive  model as the one employed for $\bl$ in \refeq{eq:priorl}:
\bee
\label{eq:prioru}
p(\mathbf{u})\propto \exp\{-\frac{1}{2} \mathbf{u}^T \mathbf{V} \mathbf{u} \}
\eee
where $\mathbf{V}=\frac{1}{\sigma_u^2} \mathbf{J}$. The matrix  $\mathbf{J}$ defined exactly as $\mathbf{H}$ in \refeq{eq:h} with proximity between two arbitrary entries $u_i$ , $u_j$ defined with respect to the nodal distance.

It is worth emphasizing that the proposed model and associated posterior contain two sets of additional parameters as compared to tradional Bayesian formulations of the inverse problem: a)  the stress vector 
$\boldsymbol{\sigma}$, and b) the model discrepancy parameters $\lambda_e^2$. The introduction of the former enables the quantification of the model discrepancy.
Despite the augmented set of parameters, these additional vectors play the role of auxiliary variables  that  expedite the exploration of the posterior using Gibbs sampling \cite{hig97aux} as it will is discussed in subsection \ref{sec:inference}. One can readily obtain, {\em conditional} posterior densities for all the parameters appearing in $\boldsymbol{\Theta}$. In particular:
\bbi
\item For $\nu^2$:
\bee
\label{eq:condnu}
\nu^{-2} \mid \mathbf{u} \sim Gamma\left(\alpha_{\nu}+\frac{n_q}{2}, \beta_{\nu}+\frac{1}{2} \parallel \mathbf{u_Q}-\mathbf{Q~u} \parallel^2 \right)
\eee
\item For $\mathbf{u}$:
\bee
\label{eq:condu}
\mathbf{u} \mid \nu^2, \boldsymbol{\sigma}, \{\mathbf{D}_e, \lambda_e^2\}_{e=1}^{n_{el}} \sim \mathcal{N}(\mathbf{\mu}_u, \mathbf{C}_u)
\eee
where:
\bee
\begin{array}{l}
 \mathbf{C}_u^{-1}=\mathbf{C}^T \boldsymbol{\Lambda}^{-1} \mathbf{C}+\frac{1}{\nu^2} \mathbf{Q}^T \mathbf{Q}+\mathbf{V} \\
\mathbf{\mu}_u=\mathbf{C}_u \left(\mathbf{C}^T \boldsymbol{\Lambda}^{-1} \boldsymbol{\sigma}+\frac{1}{\nu^2} \mathbf{Q}^T\mathbf{u}_Q \right)
\end{array}
\eee
The aforementioned matrices $\mathbf{C}$ and $\boldsymbol{\Lambda}$ arise from the model discrepancy terms in \refeq{eq:post} as follows:
\bee 
\label{eq:c}
\begin{array}{l}
\mathbf{C}=\left[ \begin{array}{l} \mathbf{D}_1 \mathbf{B}_1 \mathbf{L}_1 \\ \mathbf{D}_2 \mathbf{B}_2 \mathbf{L}_2 \\ \ldots \\ \mathbf{D}_{n_{el}} \mathbf{B}_{n_{el}} \mathbf{L}_{n_{el}} \end{array} \right] \\
\mathbf{\Lambda}=\left[ \begin{array}{llll} \lambda_1^2 \mathbf{I} & \mathbf{0} & \ldots \mathbf{0} \\ \mathbf{0} & \lambda_2^2 \mathbf{I} &  \ldots & \mathbf{\ldots} \\ \mathbf{0} & \mathbf{0} & \ldots & \lambda_{n_{el}}^2\mathbf{I}  \end{array} \right]
\end{array}
\eee
\item For $\boldsymbol{\sigma}$:
\bee
\label{eq:conds}
\boldsymbol{\sigma} \mid  \boldsymbol{u}, \{\mathbf{D}_e, \lambda_e^2\}_{e=1}^{n_{el}} \sim \mathcal{N}(\mathbf{\mu}_{\sigma}, \mathbf{C}_{\sigma})
\eee
where:
\bee
\begin{array}{l}
 \mathbf{C}_{\sigma}=\mathbf{\Lambda}+(\mathbf{\hat{B}}^T \mathbf{\Lambda})^T \left(\mathbf{\hat{B}}^T\mathbf{\Lambda} \mathbf{\hat{B}} \right)^{-1}(\mathbf{\hat{B}}^T \mathbf{\Lambda}) \\
\boldsymbol{\mu}_{\sigma}=\mathbf{C} \mathbf{u}+(\mathbf{\hat{B}}^T \mathbf{\Lambda})^T \left(\mathbf{\hat{B}}^T\mathbf{\Lambda} \mathbf{\hat{B}} \right)^{-1}(\mathbf{f}-\mathbf{\hat{B}}^T \mathbf{C}\mathbf{u})
\end{array}
\eee
\item For $\mathbf{D}_e$ assuming we are interested in the elastic modulus $E_e$ such that $\mathbf{D}_e=E_e \hat{\mathbf{D}}_e$ (where $\hat{\mathbf{D}}_e$ is known):
\bee
\label{eq:conde}
E_e \mid \boldsymbol{\sigma}_e, \mathbf{u}_e, \lambda_e^2 \sim \mathcal{N}(\mu_E,\sigma_E^2)
\eee
where:
\bee
\begin{array}{l}
 \sigma_E^{2}=\frac{\lambda_e^2}{ \parallel \hat{\mathbf{D}}_e \boldsymbol{\epsilon}_e \parallel^2 } \\
\mu_E=\frac{ \boldsymbol{\epsilon}_e^T \hat{\mathbf{D}}_e^T \boldsymbol{\sigma}_e }{\parallel \hat{\mathbf{D}}_e \boldsymbol{\epsilon}_e \parallel^2 }
\end{array}
\eee
\eei

\begin{figure}
\centering
 \psfrag{log}{ $p(\bl \mid data)$}
\psfrag{theta}{$\bl$}
\psfrag{pi}{ $p\left(\bt  \mid \bl \right)$}
\includegraphics[width=0.75\textwidth]{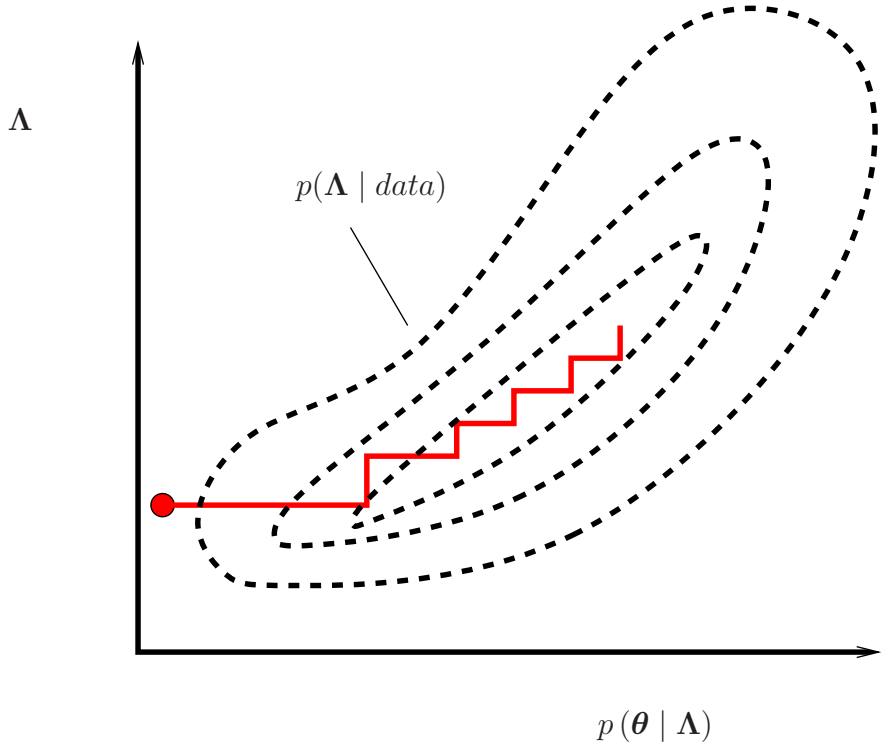}
\caption{Schematic illustration of the Expectation-Maximization scheme}
\label{fig:em}
\end{figure}

In the following we propose a hybrid scheme based on the Expectation-Maximization algorithm \cite{dem77max} that provides maximum a posteriori (MAP) point  estimates for the model discrepancy parameters $\boldsymbol {\Lambda}=\{\lambda_e^2\}$  while fully sampling from the posterior of \refeq{eq:post} for the remaining parameters $\boldsymbol{\theta}=\left(\nu^2, \boldsymbol{\sigma}, \{\mathbf{D}_e\}_{e=1}^{n_{el}}, \mathbf{u} \right)$ (Figure \ref{fig:em}).

\subsection{Inference and learning}
\label{sec:inference}

We advocate a scalable procedure for carrying out inference and learning with respect to the posterior $\pi(\mathbf{\Theta})$ (\refeq{eq:post}) which is  common practice in pertinent probabilistic models \cite{gha01int}. 
We compute point estimates for the vector  $\boldsymbol {\Lambda}=\{\lambda_e^2\}$ which  correspond to maxima $\bl^*$ of  the {\em log-posterior}.
\bee
\label{eq:logpost}
\begin{array}{ll}
L(\bl)=\log p(\mathbf{\Lambda}  \mid \mathbf{u}_Q, \mathcal{M} )& = \log \int \underbrace{p(\mathbf{\Lambda}, \boldsymbol{\theta}  \mid \mathbf{u}_Q, \mathcal{M})}_{posterior ~\refeq{eq:post} }  ~d\boldsymbol{\theta} \\
& =\log \int \pi(\mathbf{\Lambda}, \boldsymbol{\theta} ) ~d\boldsymbol{\theta}
\end{array}
\eee
while  the remaining parameters $\boldsymbol{\theta}=\left(\nu^2, \boldsymbol{\sigma}, \{\mathbf{D}_e\}_{e=1}^{n_{el}}, \mathbf{u} \right)$
are sampled from the full posterior $\pi(\bt, \bl^*)$.

 Maximization of $L(\mathbf{\bl})$ is  more complex than a standard optimization task as it involves integration over the unobserved  variables $\boldsymbol{\theta}$. 
 We propose therefore adopting an Expectation-Maximization framework (EM) which is an iterative, robust scheme that is guaranteed to increase the log-posterior at each iteration \cite{dem77max,gha01int}. 
 It is based on constructing a series of increasing lower bounds of the log-posterior using auxiliary distributions $ q(\bt)$:
\bee
\label{eq:em}
\begin{array}{ll}
L(\bt)= &= \log \int \pi(\mathbf{\Lambda}, \boldsymbol{\theta} ) ~d\boldsymbol{\theta}  \\
& = \log \int q(\bt) \frac{\pi(\mathbf{\Lambda}, \boldsymbol{\theta} ) }{q(\bt) }  ~d\bt  \\
& \ge \int q(\bt) \log \frac{\pi(\mathbf{\Lambda}, \boldsymbol{\theta} ) }{q(\bt) }  ~d\bt \quad \textrm{(Jensen's inequality)} \\
&=F(q, \bt)
\end{array}
\eee
It is obvious that this inequality becomes an equality when in place of the auxiliary distribution $q(\bt)$ the conditional posterior $\pi(\bt \mid \mathbf{\Lambda})=p(\bt \mid \mathbf{\Lambda}, \mathbf{u}_Q, \mathcal{M} )$ is selected. Given an estimate $\mathbf{\Lambda}^{(j)}$ at step  $j$, this suggests iterating between an Expectation step (E-step) whereby  we average with respect to $q^{(j)}(\bt)=\pi(\bt  \mid  \mathbf{\Lambda}^{(j)}, \mathbf{u}_Q, \mathcal{M})$ to evaluate the lower bound:
\bee
\label{eq:estep}
\begin{array}{lll}
\textrm{\bf E-step:} & F^{(j)}(q^{(j)}, \mathbf{\Lambda}) & =\int q^{(s)}(\bt) \log  \pi(\mathbf{\Lambda}, \bt )   ~d\bt  \\ & & - \int q^{(j)}(\bt ) \log  q^{(j)}(\bt ) ~d\bt
\end{array}
\eee
and a Maximization step (M-step) with respect to  $F^{(j)}(q^{(j)}, \mathbf{\Lambda})$ (and in particular the first part in \refeq{eq:estep} since the second does no not depend on $\mathbf{\Lambda}$):
\bee
\label{eq:mstep}
\begin{array}{lll}
 \textrm{\bf M-step:}  & \mathbf{\Lambda}^{(j+1)} &=arg \max_{\mathbf{\Lambda}} F^{(j)}(q^{(j)}, \mathbf{\Lambda}) \\
& & = arg \max_{\mathbf{\Theta}} E_{q^{(j)}(\bt)}\left[ \log  \pi(\mathbf{\Lambda}, \bt )  \right]  \\
& & = arg \max_{\mathbf{\Lambda}} Q(\mathbf{\Lambda}^{(j)}, \mathbf{\Lambda}) \\
\end{array}
\eee
Given the expression of the (unormalized) posterior in \refeq{eq:post}, the aforementioned objective function $Q(\mathbf{\Lambda}^{(j)}, \mathbf{\Lambda})$ becomes:
\bee
\label{eq:q}
\begin{array}{ll}
 Q(\mathbf{\Lambda}^{(j)}, \mathbf{\Lambda}) & =  E_{q^{(j)}(\bt)}\left[ \log  \pi(\mathbf{\Lambda}, \bt )  \right] \\
& = E_{q^{(j)}(\bt)}\left[ \log \prod_{e=1}^{n_{el}} p(\mathbf{c}_e \mid \boldsymbol{\sigma}_e, \mathbf{u}_e, \lambda_e^2) ~p(\boldsymbol{\Lambda})  \right] \\
& =  E_{q^{(j)}(\bt)}\left[\sum_{e=1}^{n_{el}}  \log p(\mathbf{c}_e \mid \boldsymbol{\sigma}_e, \mathbf{u}_e, \lambda_e^2) \right] + E_{q^{(j)}(\bt)}\left[ \log p(\boldsymbol{\Lambda})  \right] \\
& =\sum_{e=1}^{n_{el}} E_{q^{(j)}(\bt)}\left[\log p(\mathbf{c}_e \mid \boldsymbol{\sigma}_e, \mathbf{u}_e, \lambda_e^2) \right]+\log p(\boldsymbol{\Lambda})
\end{array}
\eee
While the second term in the expression above is essentially a penalty term arising from the prior on $\mathbf{\Lambda}$ (\refeq{eq:priorl}), the first term from \refeq{eq:pece} leads to:
\bee
\label{eq:estep2}
\begin{array}{ll}
E_{q^{(j)}(\bt)}\left[\log p(\mathbf{c}_e \mid \boldsymbol{\sigma}_e, \mathbf{u}_e, \lambda_e^2) \right]=&-\frac{n_{\sigma}}{2}\log \lambda_e^2  \\
& -\frac{1}{\lambda_e^2} E_{q^{(j)}(\bt)}\left[ \parallel 
\boldsymbol{\sigma}_e-\mathbf{D}_e \mathbf{B}_e \mathbf{u}_e \parallel^2 \right]
\end{array}
\eee
It is evident that the M-step requires computation of the sufficient statistics $\Phi_e$:
\bee
\label{eq:ss}
\Phi_e^{(j)}=E_{q^{(j)}(\bt)}\left[ \parallel 
\boldsymbol{\sigma}_e-\mathbf{D}_e \mathbf{B}_e \mathbf{u}_e \parallel^2 \right]
\eee
i.e.  the expected values (with respect to $q^{(j)}$) of the constitutive relation discrepancy in each of the elements $e=1, \ldots, n_{el}$.
Given the dependence  amongst the components of $\bl$ in the prior model, we propose an {\em incremental} version of the EM scheme (\cite{men93max,Neal}) where rather than maximizing $ Q(\mathbf{\Lambda}^{(j)}, \mathbf{\Lambda})$ in the M-step, we set $\bl^{(j+1)}$ such that:
\bee
Q(\mathbf{\Lambda}^{(j)}, \mathbf{\Lambda}^{(j+1)}) \ge Q(\mathbf{\Lambda}^{(j)}, \mathbf{\Lambda}^{(j)})
\eee 
To that end we propose maximizing $ Q(\mathbf{\Lambda}^{(j)}, \mathbf{\Lambda})$ with respect to a single component of $\bl$ (i.e. $\lambda_e^2, ~e=1, \ldots, n_{el}$) at a time while keeping the rest fixed. At each step, all the components of $\bl$ were scanned and details on the computations entailed are provided in the Appendix.

The critical task is that of inference i.e. the calculation of the expectations with respect to $q^{(j)}(\bt)$ in the E-step (\refeq{eq:estep} or \refeq{eq:estep2}). As mentioned earlier, the optimal choice for $q^{(j)}(\bt)$ is the (conditional) posterior $\pi(\bt \mid \bl^{(j)})$ which is analytically intractable as it can be readily be established from \refeq{eq:post}. While suboptimal variational approximations can be employed (e.g. \cite{gha00onl,Beal2003,wai08gra}), in this work we explore asymptotically exact, approximations based on MCMC sampling from the posterior \cite{rob04mon}. If $\{\bt^{(i,j)}\}_{i=1}^N$ denote $N$ samples from such a Markov chain with the (conditional) posterior $q^{(j)}(\bt)=\pi(\bt \mid \bl^{(j)})$ at iteration $j$ as the target, then the E-step in \refeq{eq:estep} can be substituted by:
\bee
\label{eq:mcmcem}
Q(\mathbf{\Lambda}^{(j)}, \mathbf{\Lambda})=\int q^{(j)}(\bt) \log  \pi(\mathbf{\Lambda}, \bt ) ~d\bt \approx \hat{Q}(\mathbf{\Lambda}^{(j)}, \mathbf{\Lambda})=\frac{1}{N} \sum_{i=1}^N \log  \pi(\mathbf{\Lambda}, \bt^{(i,j)} )
\eee

The unanoivadable noise introduced in these estimates by MCMC  might necessitate an exuberant   number of samples $N$ to obtain  a robust algorithm particularly close to the maximum of $L(\bl)$ (\refeq{eq:logpost}).
For that purpose we propose employing a stochastic approximation variant of the Robbins \& Monro scheme \cite{rob51sto,cap05inf}.   Rather than  increasing the simulation
size $N$ in order to reduce the variance, we compute a weighted average  at the current and previous iterations.   By employing  a decreasing sequence of weights,
information from the earlier iterations gets discarded gradually and more emphasis is placed on the
recent iterations. As it is shown in \cite{del99con}, this method converges with a fixed sample size $N$ (even when $N=1$). Convergence results that take into account the dependendence of the Markov chains at each EM-step have been obtained  by constraining the sequence of $\mathbf{\Lambda}^{(j)}$  to some compact set $\mathcal{C}$  by means of a
reprojection onto $\mathcal{C}$ \cite{kus03sto}. Even though this does not pose much problems in compuational  practice, weakened conditions  have been established in \cite{and05sta,lia07sto}.
In particular, rather than using 
 $\hat{Q}(\mathbf{\Lambda}^{(j)}, \mathbf{\Lambda})$ (which according to \refeq{eq:mcmcem} approximates ${Q}(\mathbf{\Lambda}^{(j)}, \mathbf{\Lambda})$) in the M-step (\refeq{eq:mstep}), we use:
\bee
\label{eq:saem}
\tilde{Q}(\mathbf{\Lambda}^{(j)}, \mathbf{\Lambda})=(1-\gamma_j)\tilde{Q}(\mathbf{\Lambda}^{(j-1)}, \mathbf{\Lambda})+\gamma_j \hat{Q}(\mathbf{\Lambda}^{(j)}, \mathbf{\Lambda})
\eee
where the sequence of weights $\{\gamma_j\}$ is such that   $\sum_{j=1}^\infty\gamma_j = \infty$
and $\sum_{j=1}^\infty\gamma_j^2 < \infty$ 
\footnote{A family of such sequences that was used in this work is $\gamma_j =\frac{1}{j^p}$ with  $1/2 < p \le 1$. the value of $p=0.51$ was employed}.
As it can be seen from Equations (\ref{eq:q}),   (\ref{eq:estep2}) and (\ref{eq:ss}) in order to estimate the weighted average in \refeq{eq:saem}, it suffices  to keep track of the weighted averages $\tilde{\Phi}_e^{(j)}$ of the sufficient statistics $\Phi_e^{(j)}$ (\refeq{eq:ss}):
\bee
\label{eq:saemss}
\tilde{\Phi}_e^{(j)}=(1-\gamma_j)\tilde{\Phi}_e^{(j-1)}+\gamma_j \Phi_e^{(j)}
\eee

The MCMC steps can be carried out using Gibbs sampling with respect to each of the components of $\bt$ i.e. $\nu^2$  $\mathbf{u}$, $\boldsymbol{\sigma}$ and $\{\mathbf{D}_e\}_{e=1}^{n_{el}}$ which require the conditional distributions enumerated in the previous subsection (i.e. Equations (\ref{eq:condnu}), (\ref{eq:condu}), (\ref{eq:conds}) and (\ref{eq:conde})). It is worth pointing out that  the $n\times n$ system of linear equations {\em does not need to be solved} (which has a cost of $O(n^3)$ operations) at any stage as in traditional inverse problems. If $J$ is the total number of EM iterations and $N$ is the number of MCMC steps at each iteration, then  sampling from the aforementioned conditionals implies: 
\bbi
\item the inversion and Cholesky factorization of $\mathbf{C}_u$ in order to generate samples of $\mathbf{u}$. This  must be repeated at {\em every } MCMC step since $\{ \mathbf{D}_e\}$ are updated. The cost of this operation is $O(J~ N~ n^3)$.
\item the Cholesky factorization of $\mathbf{C}_{\sigma}$ in order to generate samples of $\boldsymbol{\sigma}$. This must be repeated  at {\em every} EM iteration and {\em not at every} MCMC step since  $\mathbf{C}_{\sigma}$ solely depends on $\bl$. The cost of this operation is $O(J~(n_{el} n_{\sigma})^3)$ where where $n_{\sigma}$ is the number of stress components  ($n_{\sigma}=6$ in three dimensions,  $n_{\sigma}=3$ in plane stress/strain etc).
\eei

In order to reduce the cost associated with these operations one can employ block-Gibbs updates with respect to each of the components of $\mathbf{u}$ (or blocks of $\mathbf{u}$) rather than updating the whole vector at once.  As it is demonstrated in the sequence the cost of such a scheme is $O(J~N~n(n_{el}n_{\sigma}))$. The mixing is obviously slower than the full updates and as a consequence the variance in the MCMC estimates is larger. In general therefore  more EM iterations (assuming the same number of samples $N$ are used at each iteration) are needed to converge.  
Nevertheless the linear scaling with $J$ constitutes such a scheme more efficient. Similar block-Gibbs updates can be carried out for $\boldsymbol{\sigma}$ reducing the cost associated with this task to $J~N~n(n_{el}n_{\sigma}))$.
The conditional posteriors for performing block-Gibbs moves are described in the sequence.

Let $\left[ \begin{array}{l}u_i \\ \mathbf{u}_{-i} \end{array} \right]$ be a partitioning of $\mathbf{u}$ with respect to component $i$\footnote{An identical procedure can be followed when $u_i$ corresponds to a block of $\mathbf{u}$}.  Let also  $\mathbf{Q}=\left[  \mathbf{Q}_i \mid \mathbf{Q}_{-i} \right]$, $\mathbf{C}=\left[ \mathbf{C}_i \mid \mathbf{C}_{-i} \right]$ the corresponding partitioning of the matrices appearing in \refeq{eq:disp} and \refeq{eq:c}. Then the conditional posterior of $u_i$ from \refeq{eq:post} is:
\bee
\label{eq:condui}
u_i \mid \mathbf{u}_{-i}, \nu^2, \boldsymbol{\sigma}, \{\mathbf{D}_e, \lambda_e^2\}_{e=1}^{n_{el}} \sim \mathcal{N}(\mu_{u_i}, \sigma_{u_i}^2)
\eee
where:
\bee
\sigma_{u_i}^{-2}=\mathbf{C}_i^T \boldsymbol{\Lambda}^{-1} \mathbf{C}_i+\frac{1}{\nu^2} \mathbf{Q}_i^T \mathbf{Q}_i 
\eee
\bee                                                                                                                   
\mu_{u_i}=\sigma_{u_i}^2 \left(\mathbf{C}_i^T \boldsymbol{\Lambda}^{-1} (\boldsymbol{\sigma}-\mathbf{C}_{-i}\mathbf{u}_{-i})+\frac{1}{\nu^2} \mathbf{Q}_i^T(\mathbf{u}_Q-\mathbf{Q}_{-i}\mathbf{u}_{-i}) \right)
\eee
%
%

It is noted that the leading order of computational operations for updating succesively all components of $\mathbf{u}$ as above is $O(n (n_{el}n_{\sigma}))$. This is approximately one order  less than the $O(n^3)$ cost associated with the full update (\refeq{eq:condu}), considering that the dimension of the stress vector $n_{el}n_{\sigma}$ is comparable to $n$.

\section{Numerical examples}
\label{sec:examples}

\begin{figure}
\begin{minipage}[b]{0.45\linewidth}
\centering
\psfrag{u1}{$u_1=1$}
\psfrag{u2}{$u_2=1$}
 \includegraphics[width=0.95\textwidth]{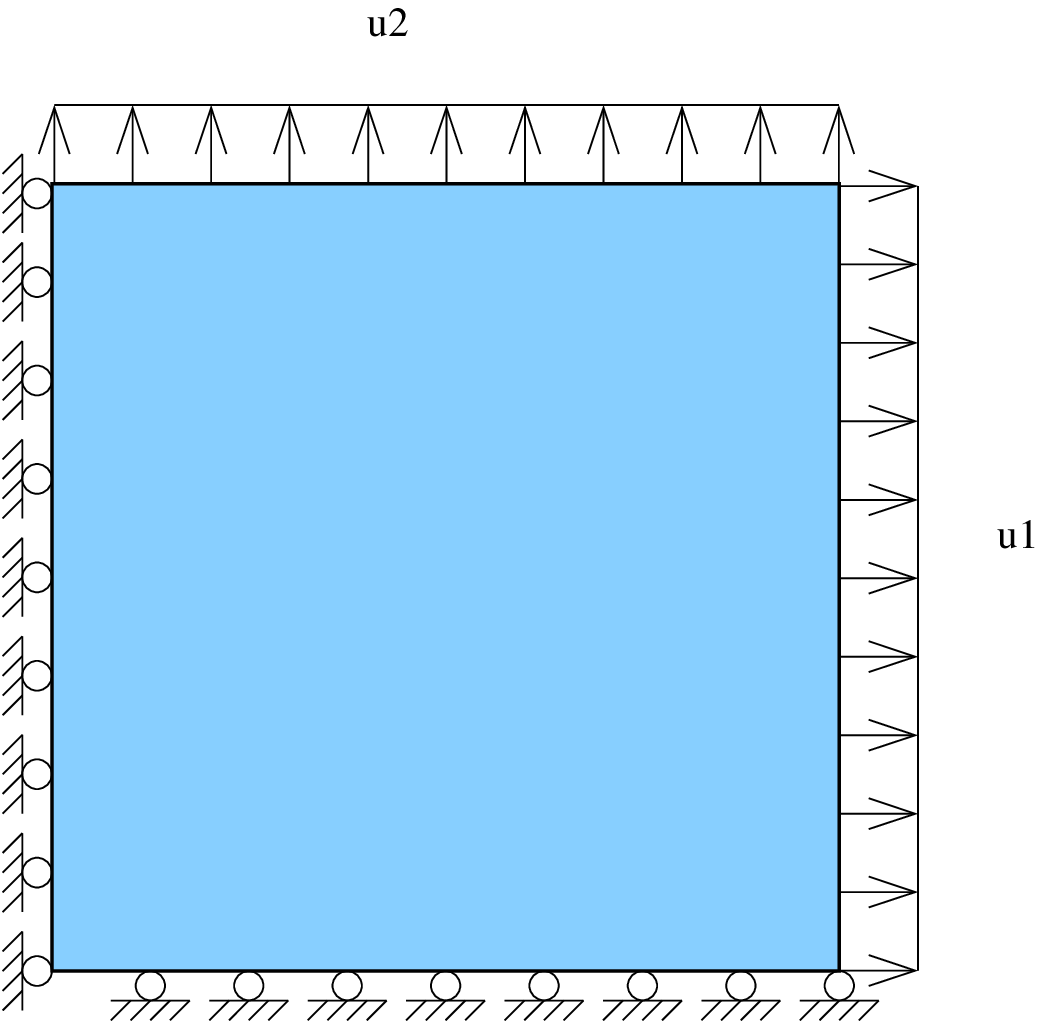}
\caption{ Problem configuration used in both examples 1 and 2 \protect\cite{alb09adj}  }
\label{fig:conf}
\end{minipage}
\hfill
\begin{minipage}[b]{0.45\linewidth}
\centering
\raisebox{-1.5cm}{\includegraphics[width=0.95\textwidth]{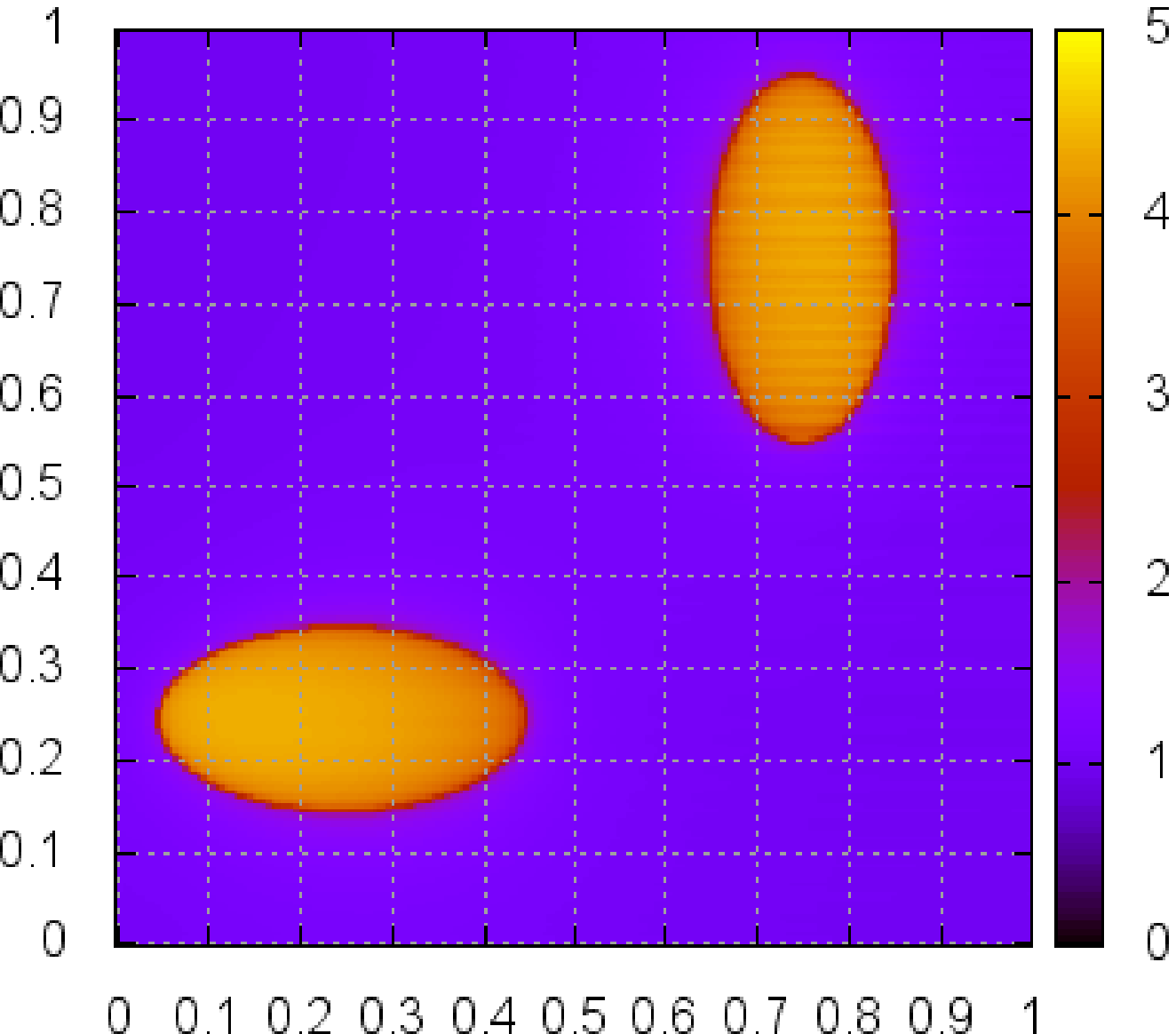}}
\caption{Example 1 - Elastic modulus $E$ spatial distribution:  In the inclusions $E=5$ whereas in the rest of the domain \protect$E=1$}
\label{fig:two}
\end{minipage}
\end{figure}

In this section we report results on the accuracy and performance of the algorithm on two-dimensional elastography problems on synthetic data obtained for the configuration depicted in Figure \ref{fig:conf} \cite{alb09adj,ISI:000275699600006} where the  boundary displacements  normal to the walls are prescribed .
  We intend to provide a clinical validation of the approach in a future study.

We assume an isotropic  elastic material with  Poisson's ratio  $\nu=0.5$ (incompressible) 
and employ the selective  reduced integration quadrilateral elements for the forward problem \cite{hug80gen,hug00fin}.
%
We examine two distributions for the elastic  modulus occurring in
 elliptic  and circular inclusions. In the first problem (Figure \ref{fig:two}) the emphasis is on demonstrating the capabilities of the proposed method in identifying the ground truth as well as providing probabilistic confidence metrics particularly in the presence of noise. In the second case (Figure \ref{fig:one}) the emphasis is on detecting and quantifying model discrepancies in the sense described in section \ref{sec:meth}. It is noted that in all cases apart from the identification of material properties, a  direct output of the computations is the stress distribution.
It is finally noted that in order to generate the displacement  data, the forward problem was solved with a randomly generated mesh consisting of $10,000$  elements.

The following values were used for the parameters appearing in prior models described previously:
\begin{itemize}
\item $\sigma_z^2=100$ (\refeq{eq:priorl}) which corresponds to a diffuse prior and $\sigma_u^2=1$ in \refeq{eq:prioru}. The latter was selected based on the magnitude of the prescribed boundary displacements in Figure \ref{fig:conf}.
\item $d_0=0.1$ for the correlation-length parameter appearing in the $\mathbf{H}$ (\refeq{eq:h}) and  $\mathbf{V}$ (\refeq{eq:prioru}). Numerical evidence suggested that the effect of this parameter was minimal when varied in the range $[0.01,0.5]$ given that the characteristic dimension of the problem domain is $1$.
\item an uniformative Jeffry's prior was adopted for the observation noise variance $\nu^2$ (\refeq{eq:condnu}) with $\alpha_{\nu}=2$ and $\beta_{\nu}=0$.
\end{itemize}

With regards to the EM scheme, at each iteration $N=10$ MCMC updates of all model parameters were performed and iterations were terminated when the {\em relative increase} in the objective $Q(\bl^{(j)},\bl)$ in \refeq{eq:q} was less or equal than $\epsilon=0.001$, i.e. $\frac{ \mid Q(\bl^{(j)},\bl)-Q(\bl^{(j-1)},\bl) \mid }{\mid Q(\bl^{(j-1)},\bl) \mid } <\epsilon$.

\subsection{Example 1}

The first scenario involves two elliptical inclusions centered at $(0.25,0.25)$ and $(0.75,0.75)$ with principal axes $0.1$ and $0.2$.
 with a contrast ratio $5:1$ in the elastic modulus (Figure \ref{fig:two}). A useful outcome of the numerical investigations was the fact that the overall inference and learning process can be greatly accelerated by operating on a sequence of discretizations with increased refinement. In particular, initially a coarse mesh is adopted with few nodes and elements where the proposed EM scheme is applied. The parameter values learned (i.e. $\bl$) are used as the initial values for a refined mesh. The MCMC chains with respect to the other model parameters at the new mesh are initiated from samples drawn at the coarser mesh. It was found that this led to a reduction of the number of EM iterations needed to achieve convergence and significant acceleration since the order of operations at coarse meshes is smaller.
For that purpose, we report in this problem the results obtained at three different resolutions employing a regular mesh with $5\times 5$ , $10\times 10$ and $20 \times 20$ elements.
A potentially important implication involves the possibility of  {\em adaptive refinement} where the mesh can be refined at selected regions of the problem domain where further information is needed as determined by the inferences produced at coarser resolutions \cite{ISI:000275756200016}.

\begin{figure}[!h]
%
\subfigure[20x20: $5\%$ quantile]{
\includegraphics[width=0.3\textwidth]{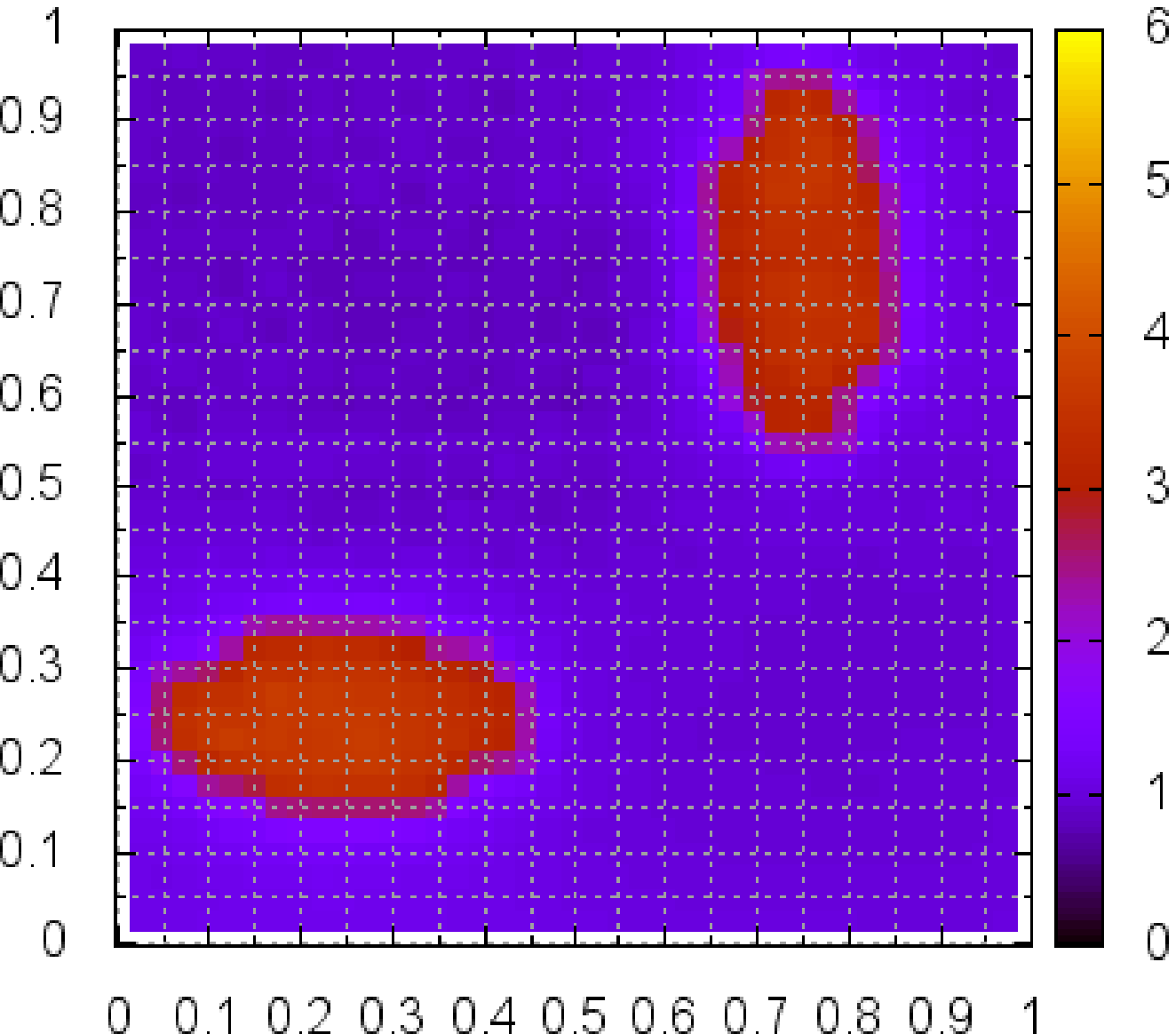}}
\hfill
\subfigure[20x20: posterior mean]{
\includegraphics[width=0.3\textwidth]{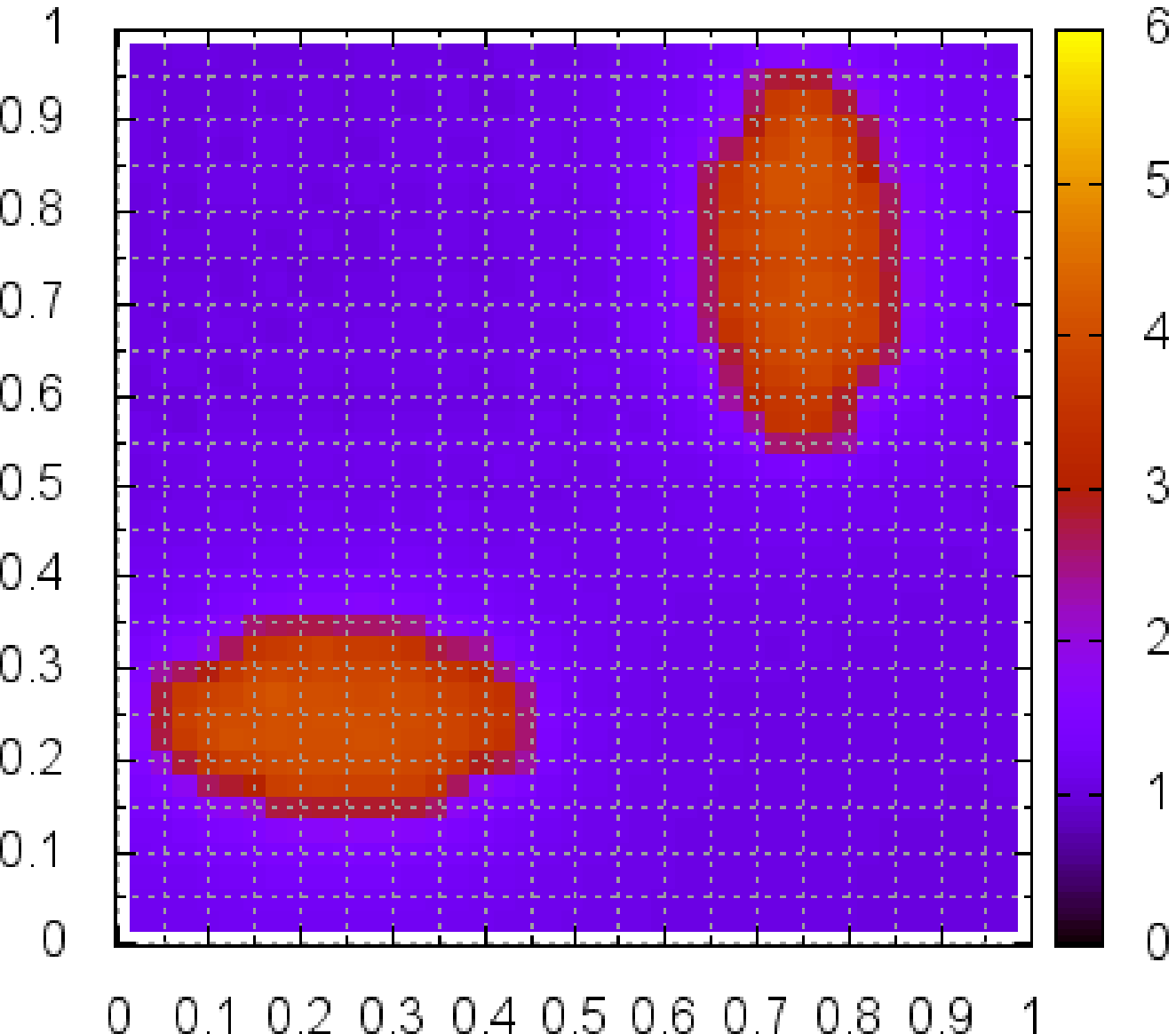}}
\hfill
\subfigure[20x20: $95\%$ quantile]{
\includegraphics[width=0.3\textwidth]{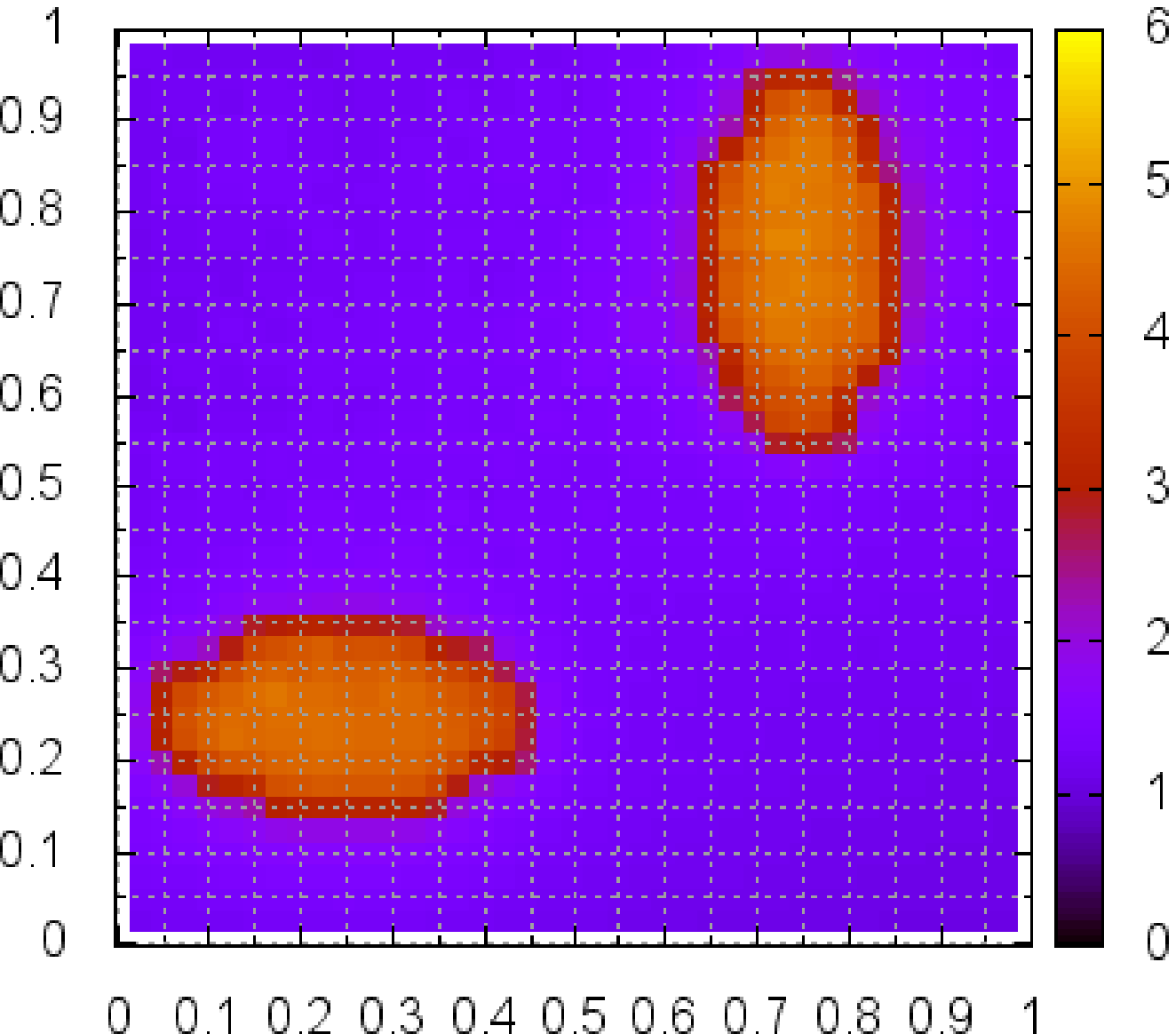}}
\caption{Example 1: Posterior statistics of the elastic modulus distribution for noiseless data.}
\label{fig:two_ellipses_snr=0}
\end{figure}

\begin{figure}[!h]
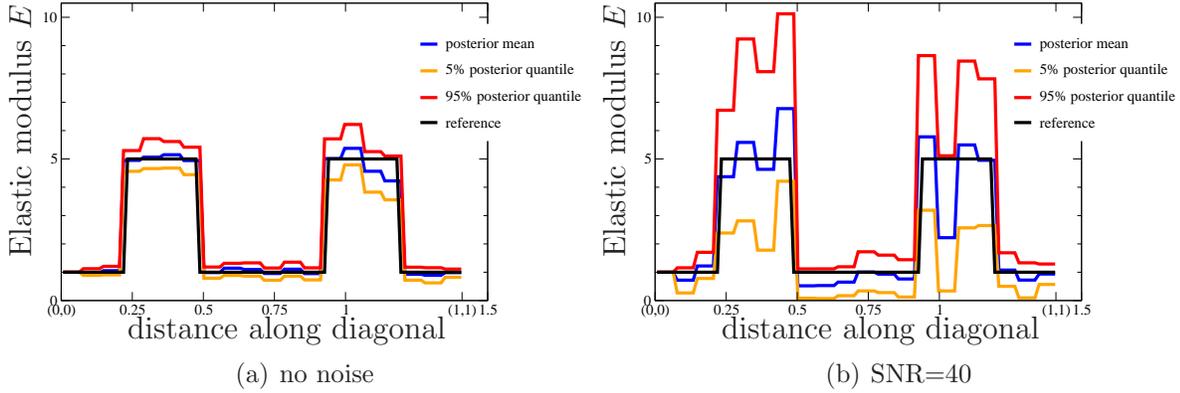

 \vspace{.5cm}
\psfrag{distance}{distance along diagonal}
\psfrag{modulus}{Elastic modulus $E$}
\subfigure[no noise]{
\includegraphics[width=0.48\textwidth]{FIGURES/two_ellipses_diag_snr=0.eps}
\label{diaga}}
\hfill
\subfigure[SNR=40]{
\includegraphics[width=0.48\textwidth]{FIGURES/two_ellipses_diag_snr=40.eps}
\label{diagb}
}
\caption{Example 1: Posterior statistics of the elastic modulus distribution along the diagonal from $(0,0)$ to $(1,1)$.}
 \label{fig:two_ellipses_diag}
\end{figure}

Figure \ref{fig:two_ellipses_snr=0} depicts the posterior mean as well as the posterior quantiles at $5\%$ and $95\%$ for the elastic modulus at these three resolutions and in the absence of noise in the data. It is readily observed that the proposed scheme can identify the ground truth as well as provide posterior credible intervals on the inferences made. These are more clearly depicted in Figure \ref{diaga} which presents the results along the diagonal from $(0,0)$ to $(1,1)$.

\begin{figure}[!h]
%
\subfigure[20x20: $5\%$ quantile]{
\includegraphics[width=0.3\textwidth]{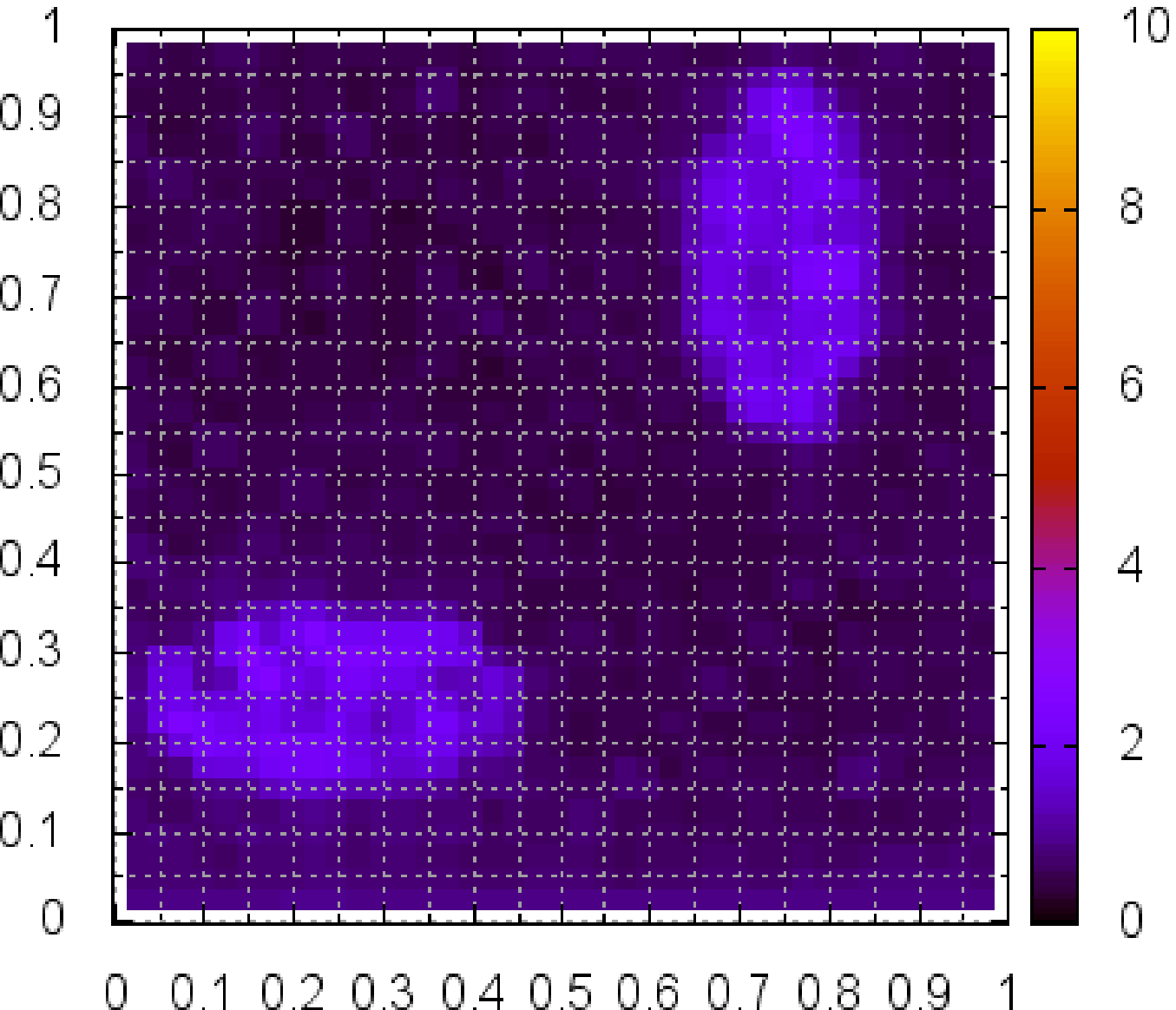}}
\hfill
\subfigure[20x20: posterior mean]{
\includegraphics[width=0.3\textwidth]{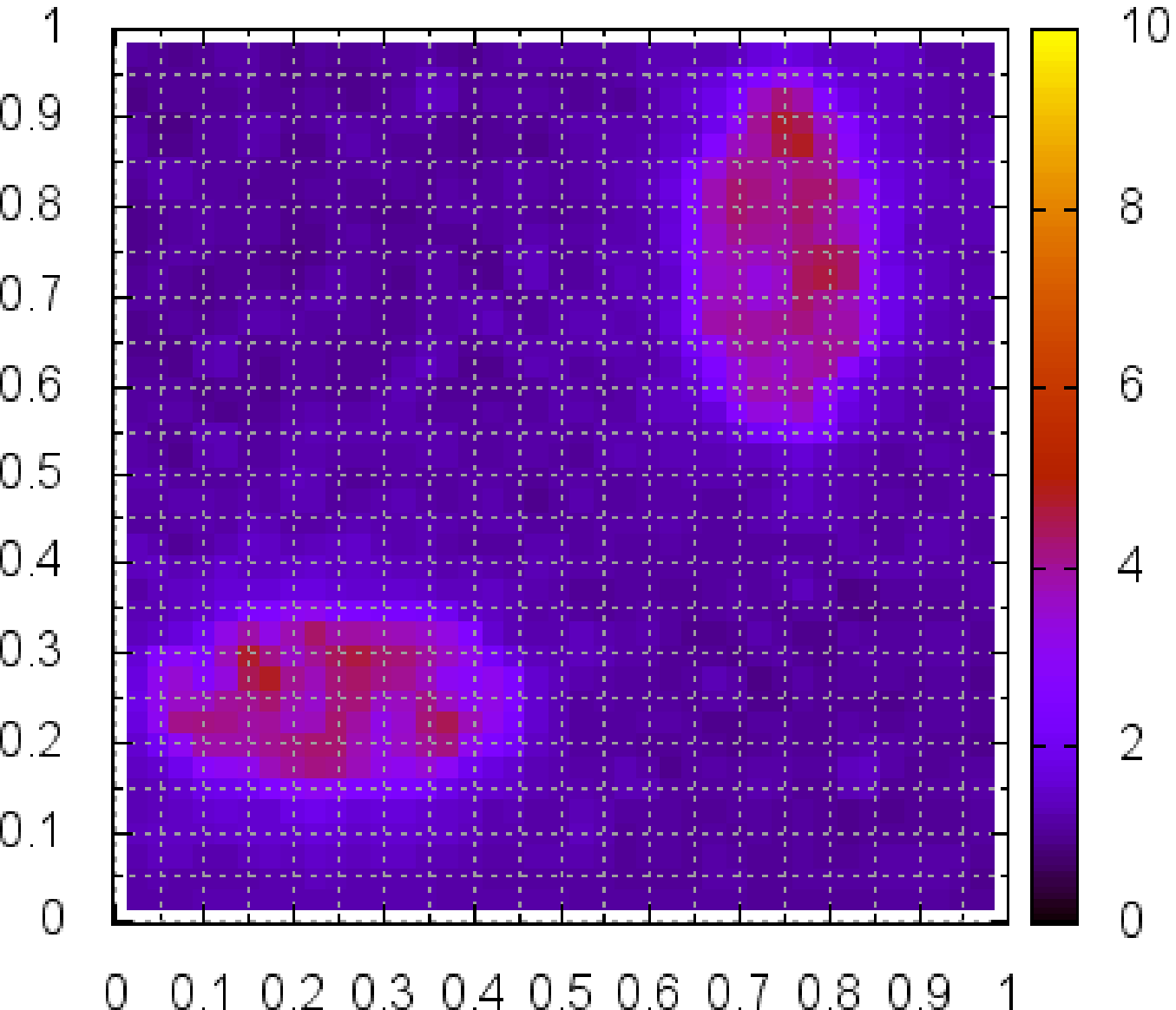}}
\hfill
\subfigure[20x20: $95\%$ quantile]{
\includegraphics[width=0.3\textwidth]{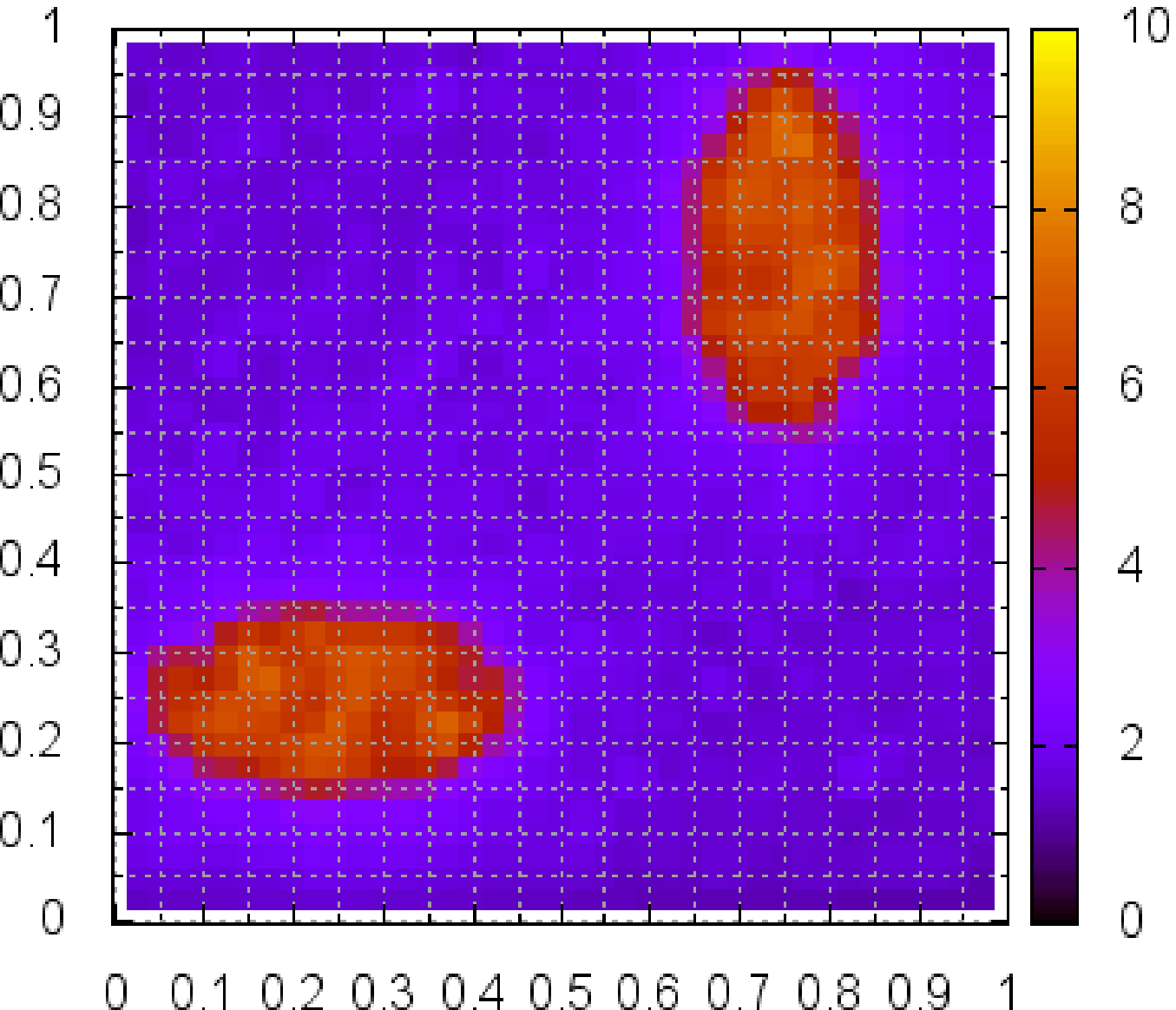}}
\caption{Example 1: Posterior statistics of the elastic modulus distribution for noisy data with SNR=$40dB$.}
\label{fig:two_ellipses_snr=40}
\vspace{.5cm}
\end{figure}

 We also investigated the performance of the algorithm in the presence of zero mean, Gaussian noise and in particular with a Signal-to-Noise-Ratio $SNR=40dB$ which is typical for ultrasound systems \cite{lakisorig,ISI:000223500200013}. The results are shown in  Figure \ref{fig:two_ellipses_snr=40} in terms of posterior mean and posterior quantiles. As it can also be seen in Figure \ref{diagb}, the algorithm is able to quantify the uncertainty introduced by the presence of noise and posterior bounds provided enclose the ground truth.
Finally Figure \ref{fig:two_ellipses_samples} depicts randomly selected samples drawn at various iterations of the EM scheme (for the finest resolution $20\times 20$) that demonstrate the evolution of the learning algorithm proposed.

%
%

\subsection{Example 2}

\begin{figure}
\centering
\includegraphics[width=0.5\textwidth]{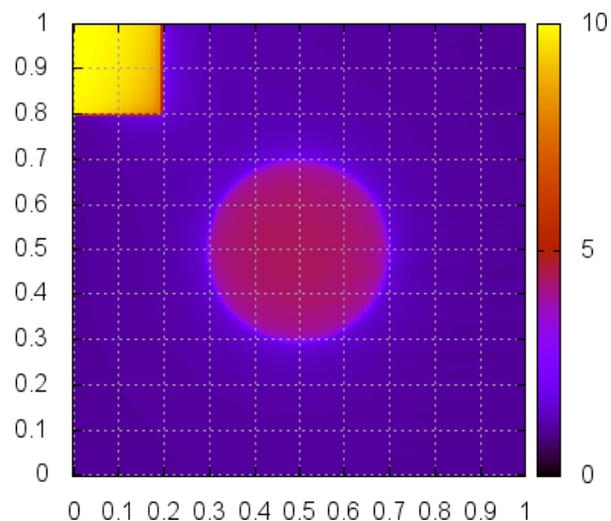}
\caption{Example 2 - Elastic modulus $E$ spatial distribution}
\label{fig:one}
\end{figure}

The primary goal in the second example is to demonstrate the capability of   quantifying model discrepancy in the constitutive equation. In particular we consider the synthetic data generated by the material distribution in Figure \ref{fig:one} \footnote{In the circular inclusion $E=5$, in the subdomain  $[0, 0.2]\times [0.8,1]$  we employed a constitutive matrix \protect$\mathbf{D}= \left[ \begin{array}{ccc} 10 & -5 & -5 \\  -5, & 10. &  -5 \\    -5 & -5 &  100   \end{array} \right]$ whereas in the rest of the domain \protect$E=1$}.
 The circular inclusion centered at $(0.5, 0.5)$ with radius $0.2$ is assumed to have an elastic modulus that is $5$ times larger than the rest of domain. We further assumed a square region on the top left corner $[0, 0.2]\times [0.8,1]$ where rather than an {\em isotropic}, elastic material we employed an anisotropic constitutive matrix $\mathbf{D}=\left[ \begin{array}{ccc} 10 & -5 & -5 \\  -5, & 20. &  -5 \\    -5 & -5 &  100  \end{array} \right]$.
While this is a valid constitutive model (i.e. $\mathbf{D}$ is positive definite) it is obviously inconsistent with the isotropic assumption made in the model used to identify material properties.
While other inversion schemes might be able to  find an elastic modulus corresponding to an isotropic material that  fits adequately the observed displacemnts, they would be unable to identify that the model employed is inadequate. As a result  erroneous conclusions would be drawn  about the state of the material at this portion of the problem domain.

\begin{figure}[!h]
\subfigure[no noise]{
\includegraphics[width=0.48\textwidth]{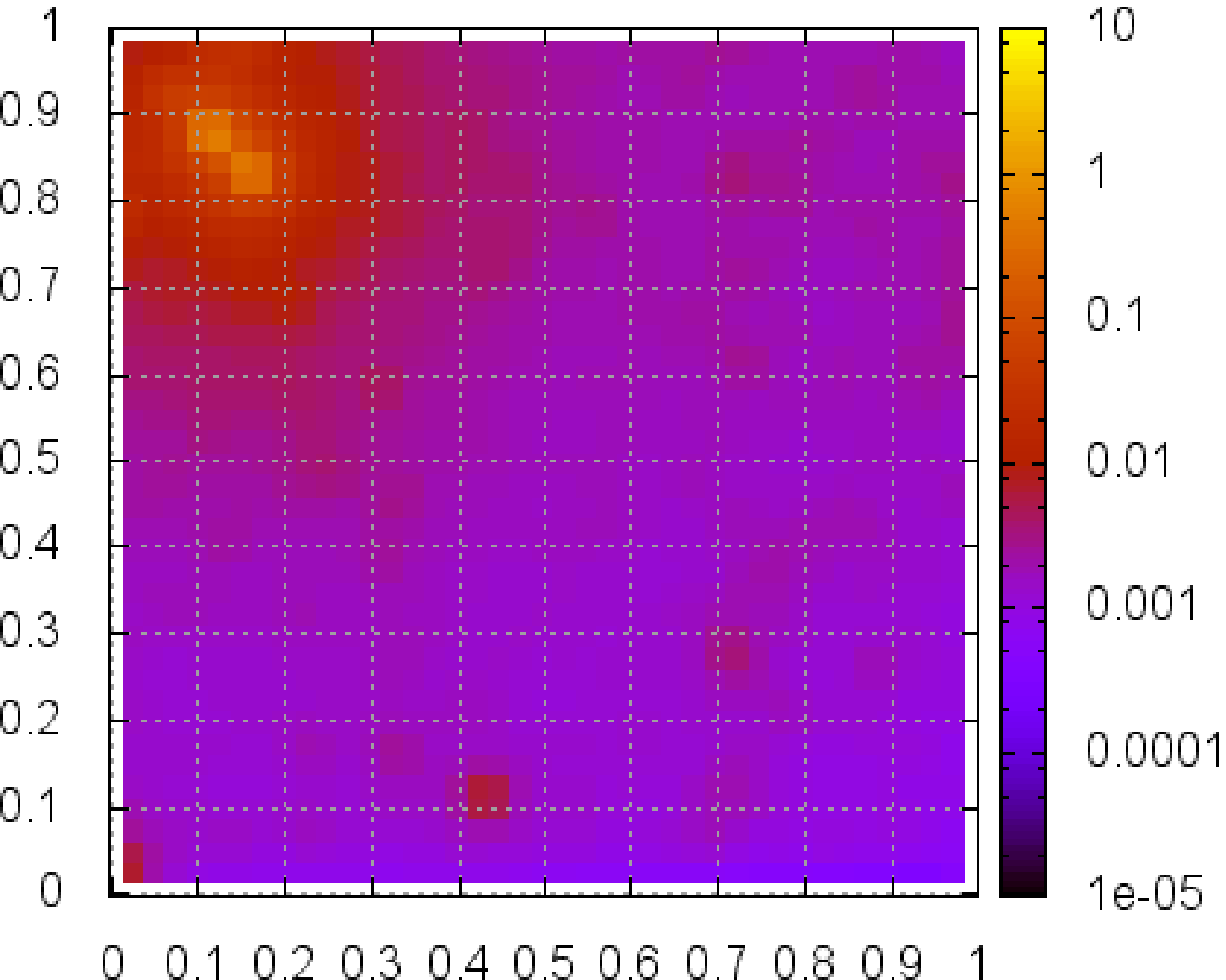}
\label{sl2a}}
\hfill
\subfigure[SNR=40]{
\includegraphics[width=0.48\textwidth]{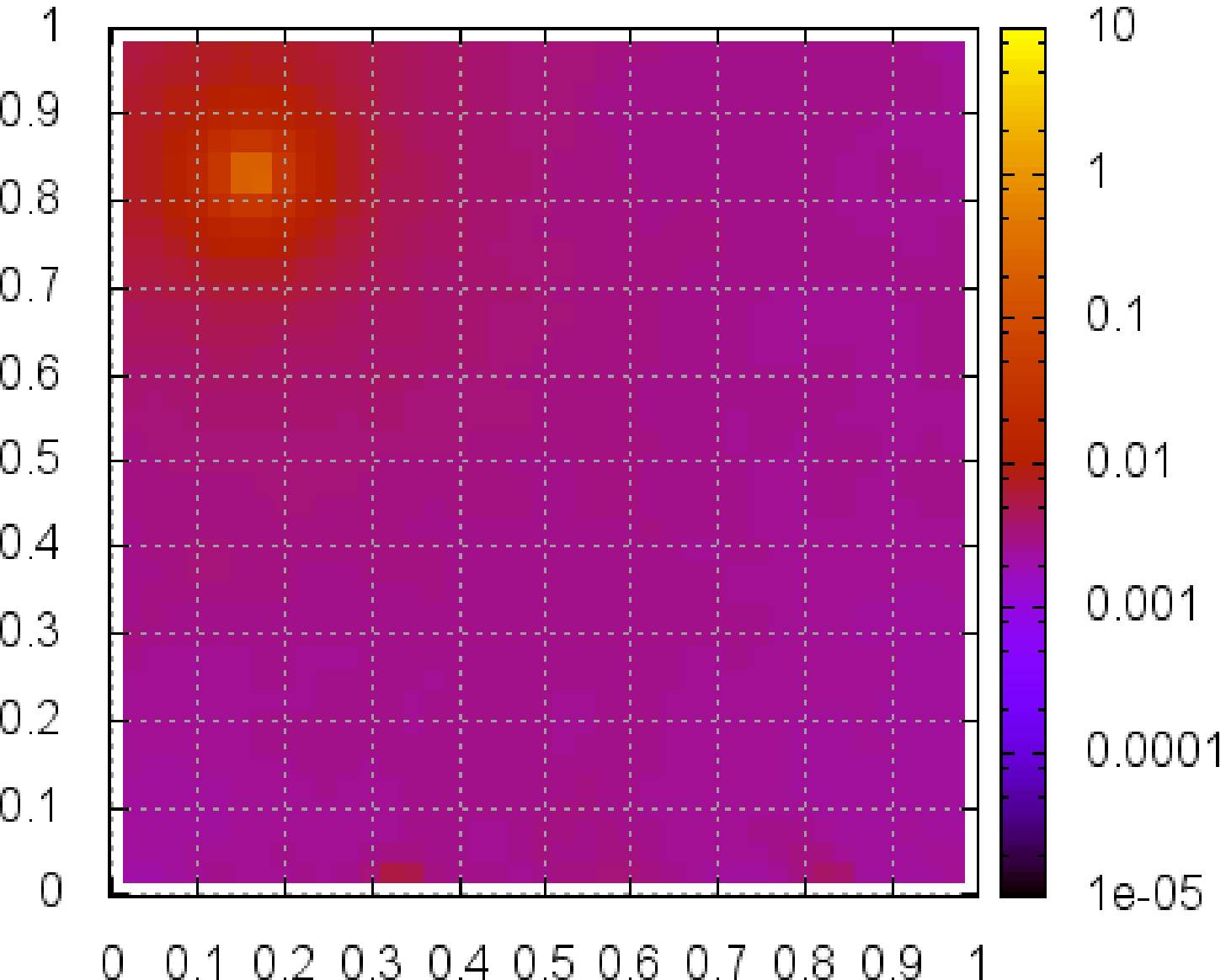}
\label{sl2b}
}
\caption{Example 2: Model discrepancies/errors $\{ \lambda_e^2\}_{e=1}^{n_{el}}$ for a)  no noise, and b) SNR=$40 dB$ (in log-scale)}
 \label{fig:one_ellipse_sl2}
\end{figure}

\begin{figure}[!h]
\subfigure[no noise - $5\%$ quantile]{
\includegraphics[width=0.3\textwidth]{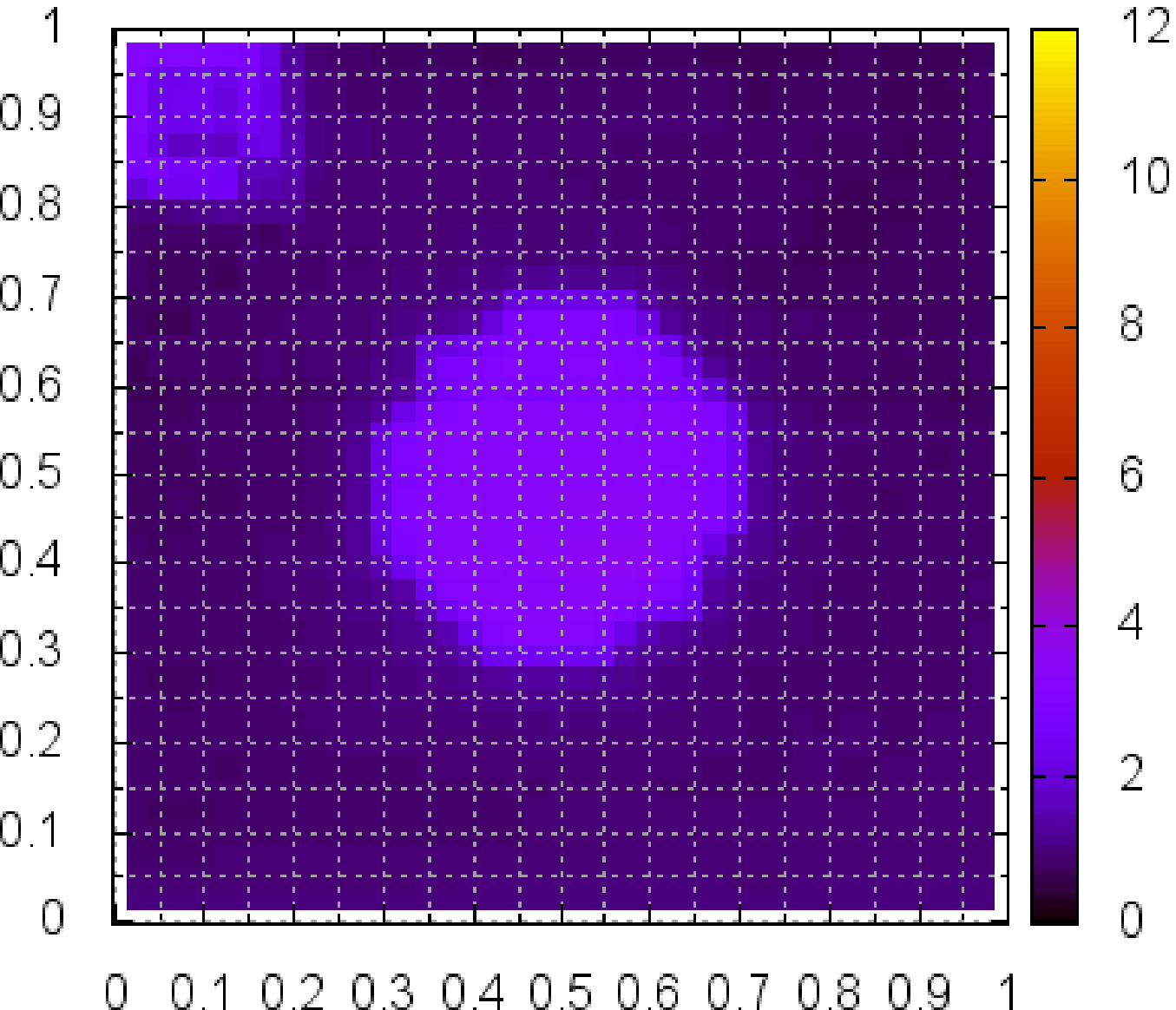}}
\hfill
\subfigure[no noise - posterior mean]{
\includegraphics[width=0.3\textwidth]{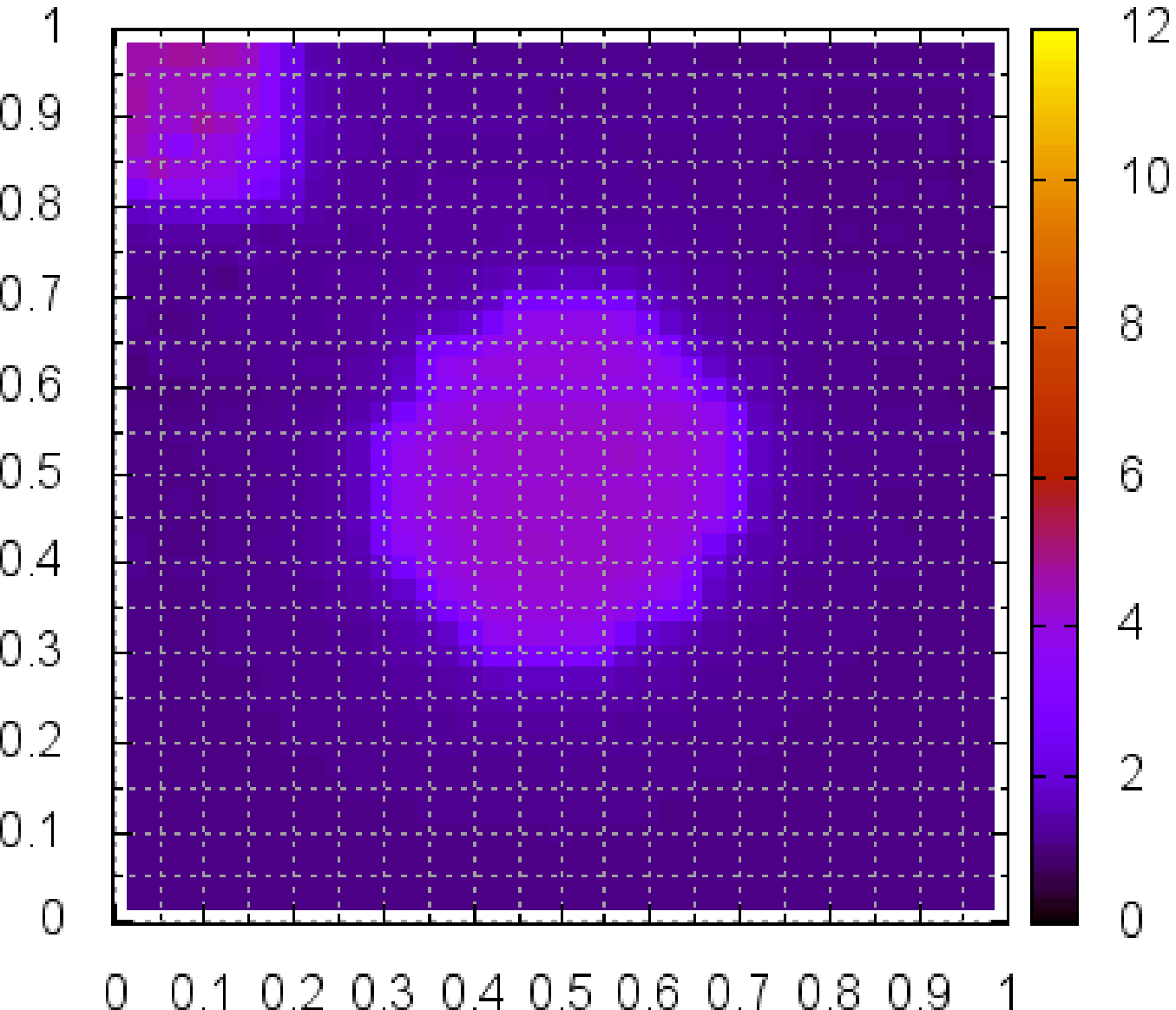}}
\hfill
\subfigure[no noise -  $95\%$ quantile]{
\includegraphics[width=0.3\textwidth]{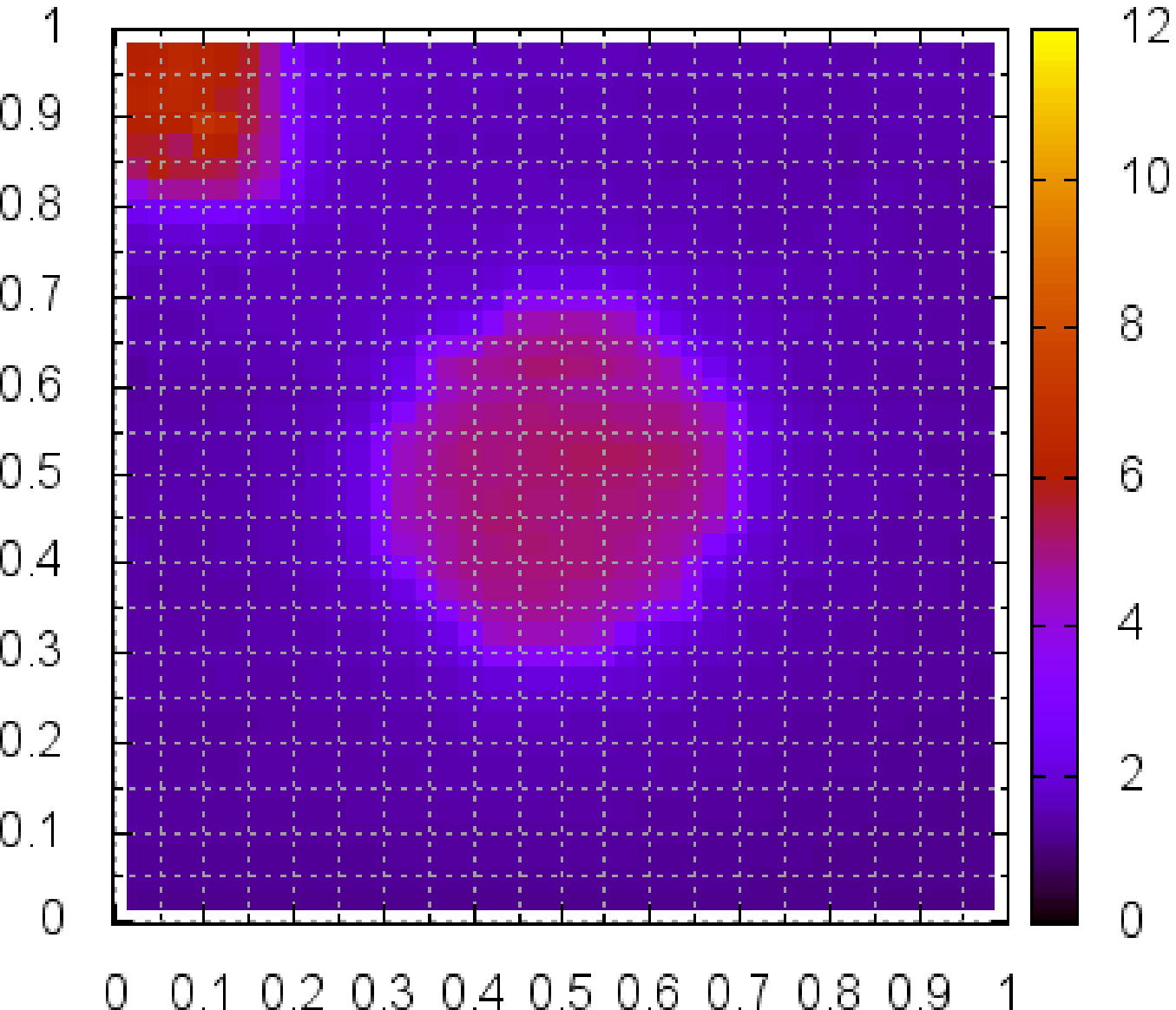}}
\\
\subfigure[SNR=40 - $5\%$ quantile]{
\includegraphics[width=0.3\textwidth]{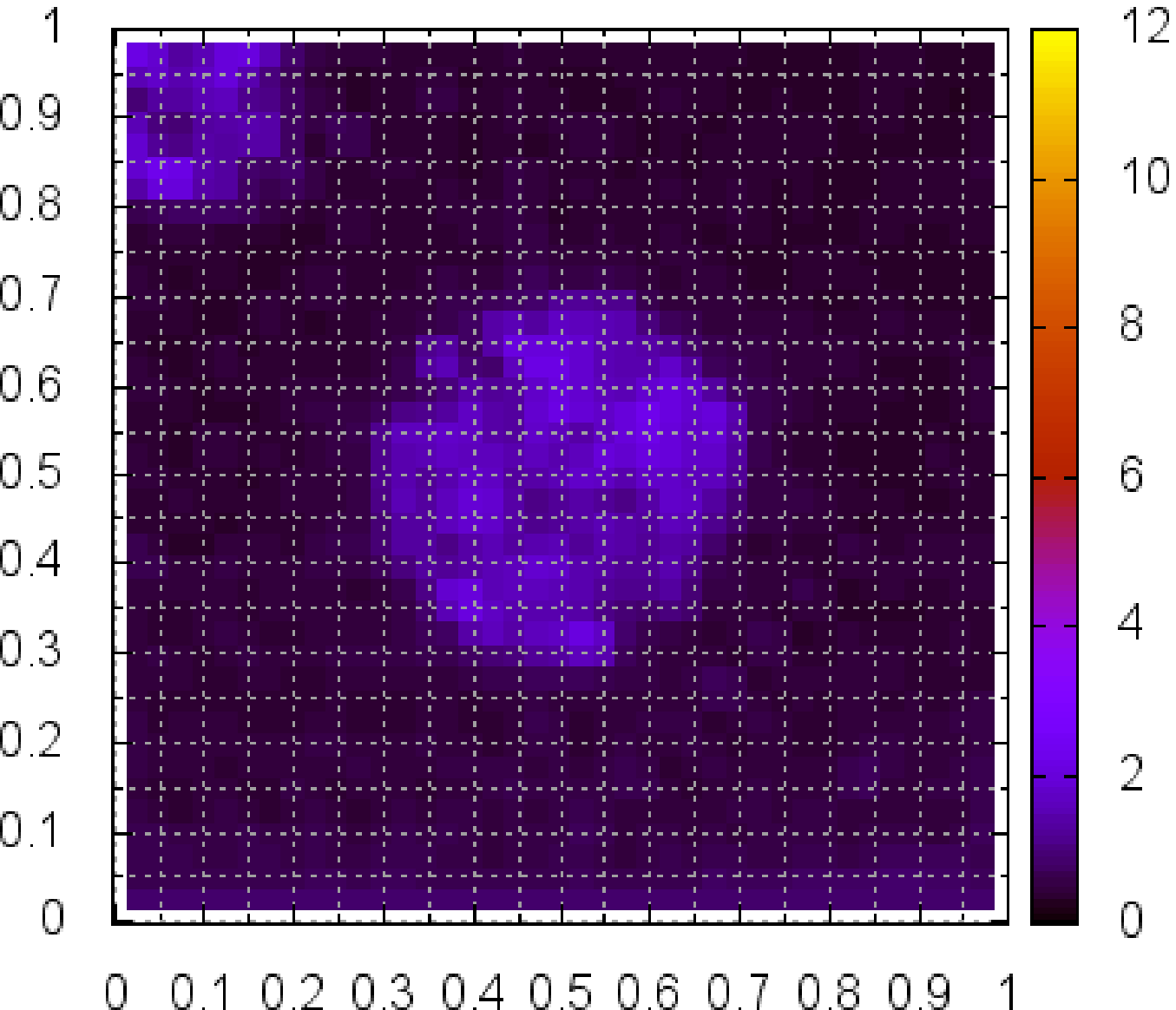}}
\hfill
\subfigure[SNR=40 - posterior mean]{
\includegraphics[width=0.3\textwidth]{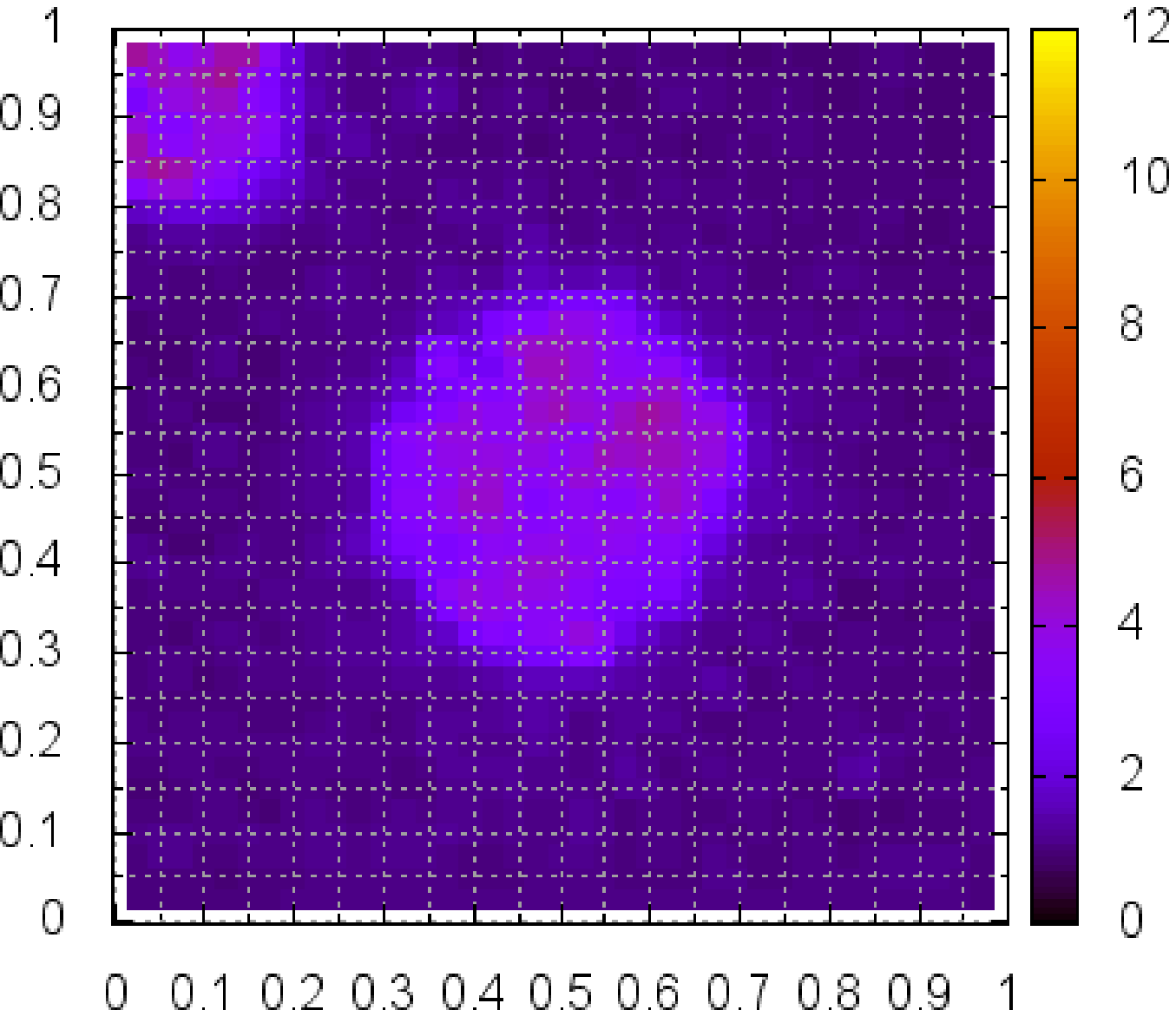}}
\hfill
\subfigure[SNR=40 -  $95\%$ quantile]{
\includegraphics[width=0.3\textwidth]{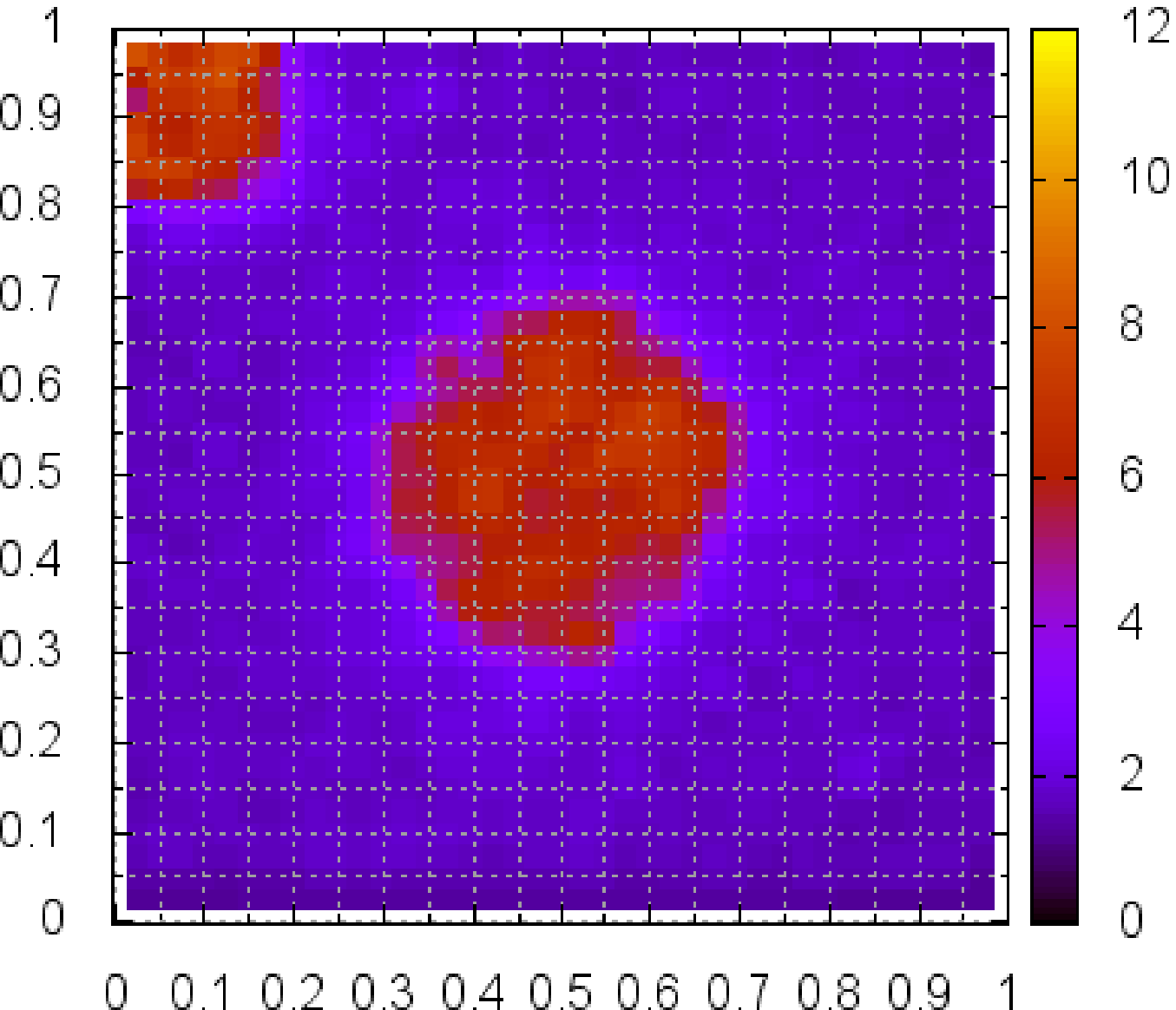}}
\caption{Example 2: Posterior statistics of the elastic modulus distribution when data have no noise and for SNR=$40dB$.}
\label{fig:one_ellipse} 
\end{figure}

\begin{figure}[!h]
\subfigure[reference (ground truth)]{
\includegraphics[width=0.3\textwidth]{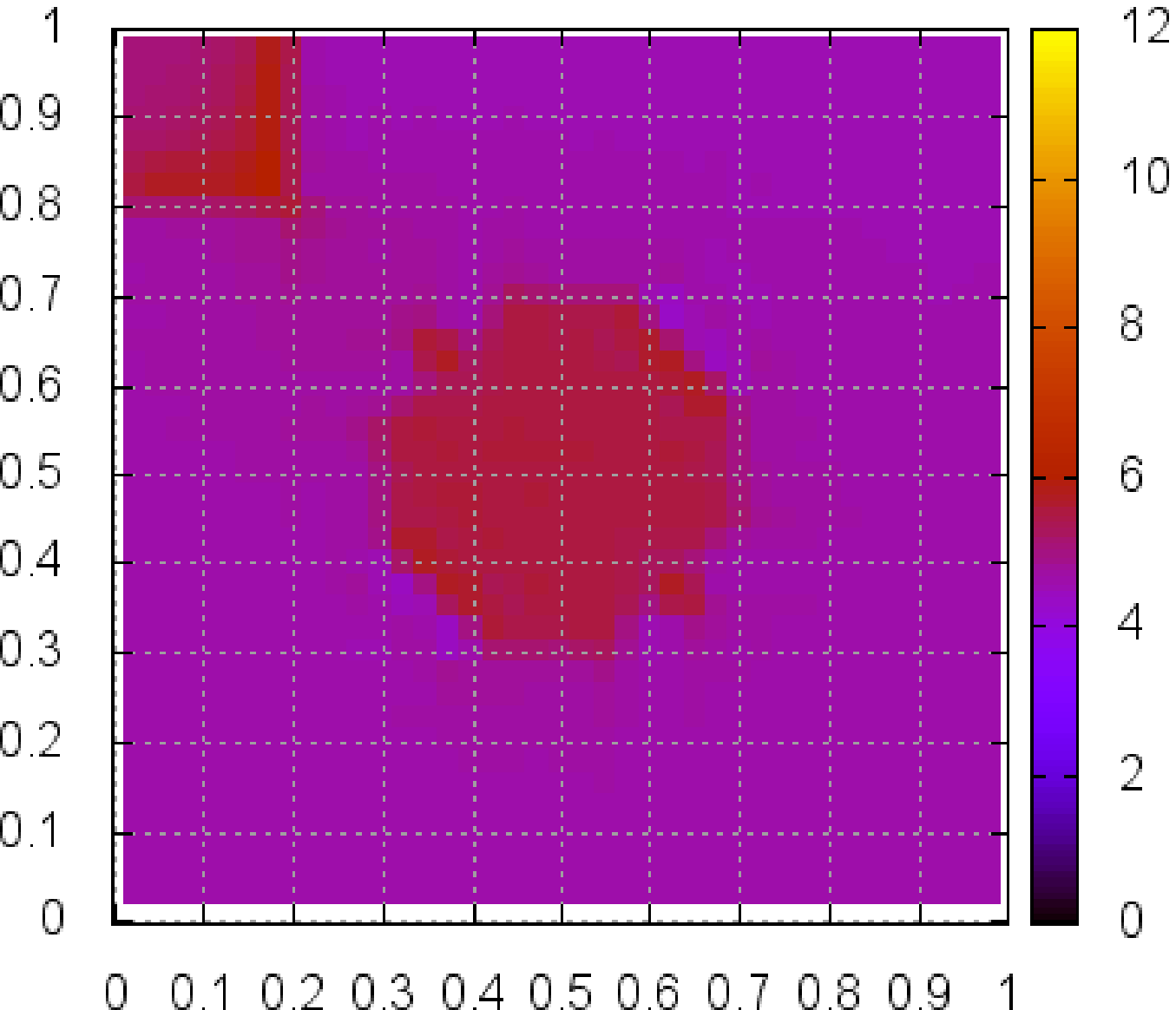}}
\hfill
\subfigure[no noise - posterior mean]{
\includegraphics[width=0.3\textwidth]{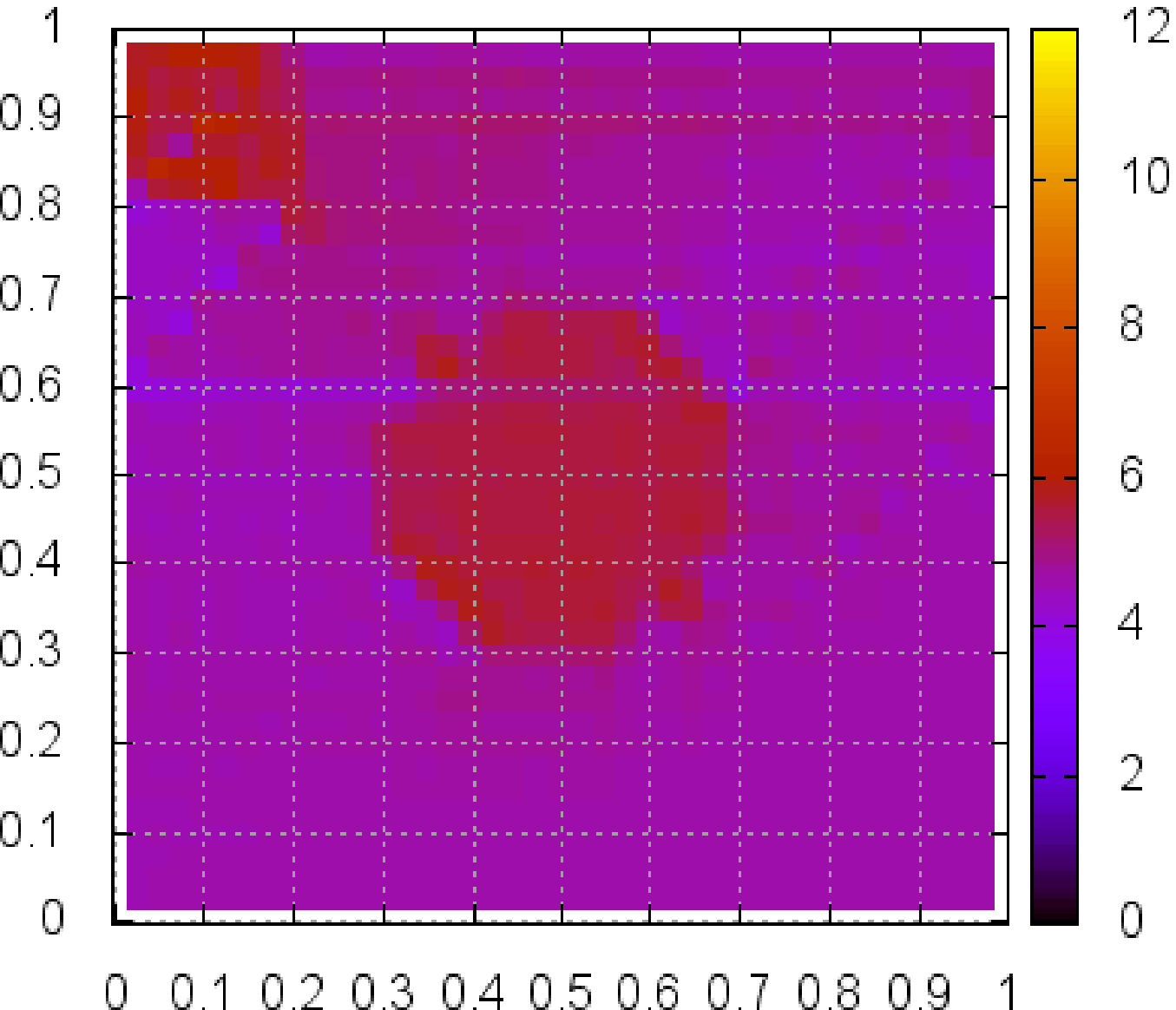}}
\hfill
\subfigure[SNR=40 - posterior mean]{
\includegraphics[width=0.3\textwidth]{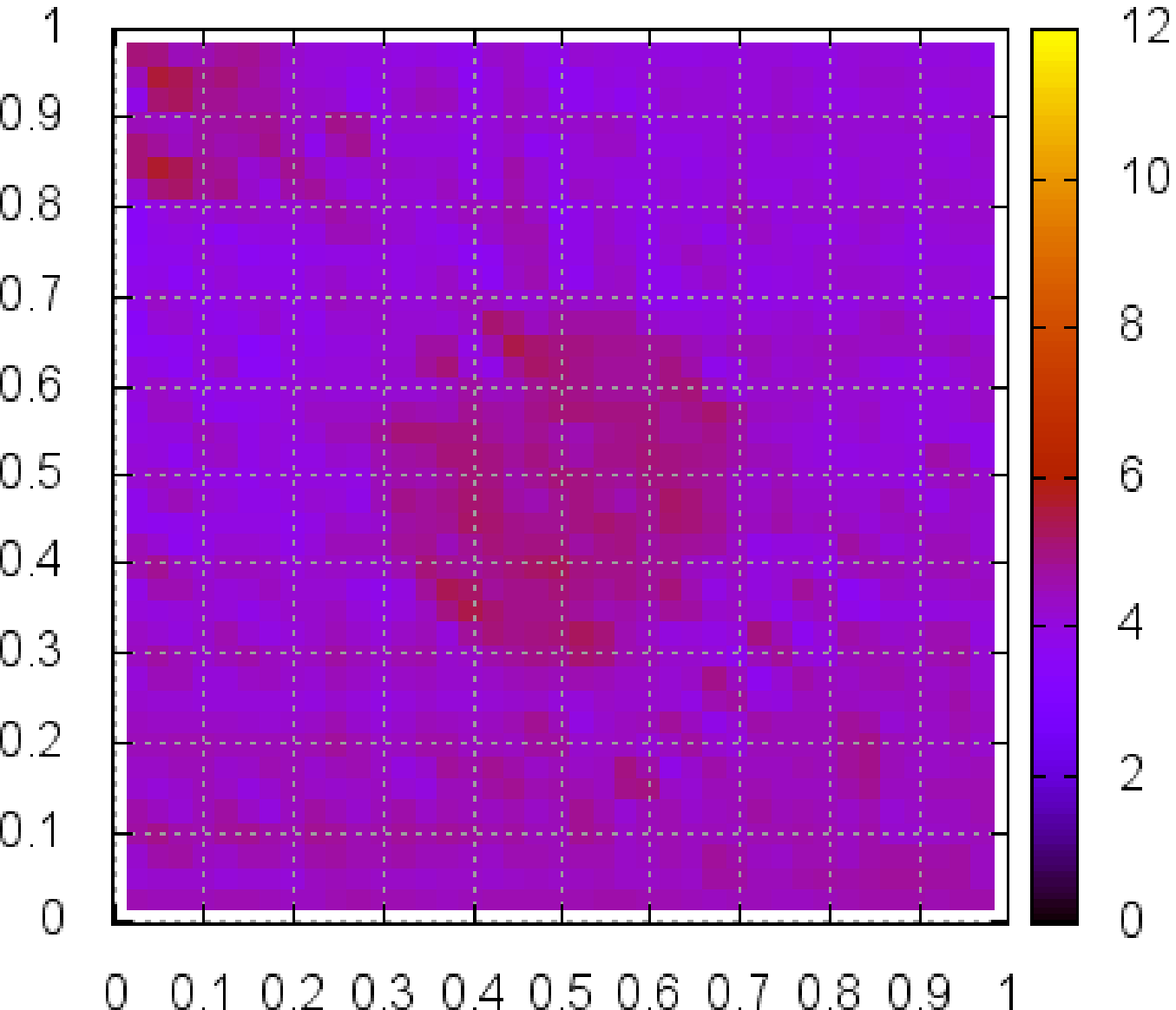}}
\caption{Example 2: Comparison of pressure's spatial distribution with the posterior means obtained when data have no noise and for SNR=$40dB$.}
\label{fig:one_ellipse_pressure} 
\end{figure}

\begin{figure}[!h]
\subfigure[reference (ground truth)]{
\includegraphics[width=0.3\textwidth]{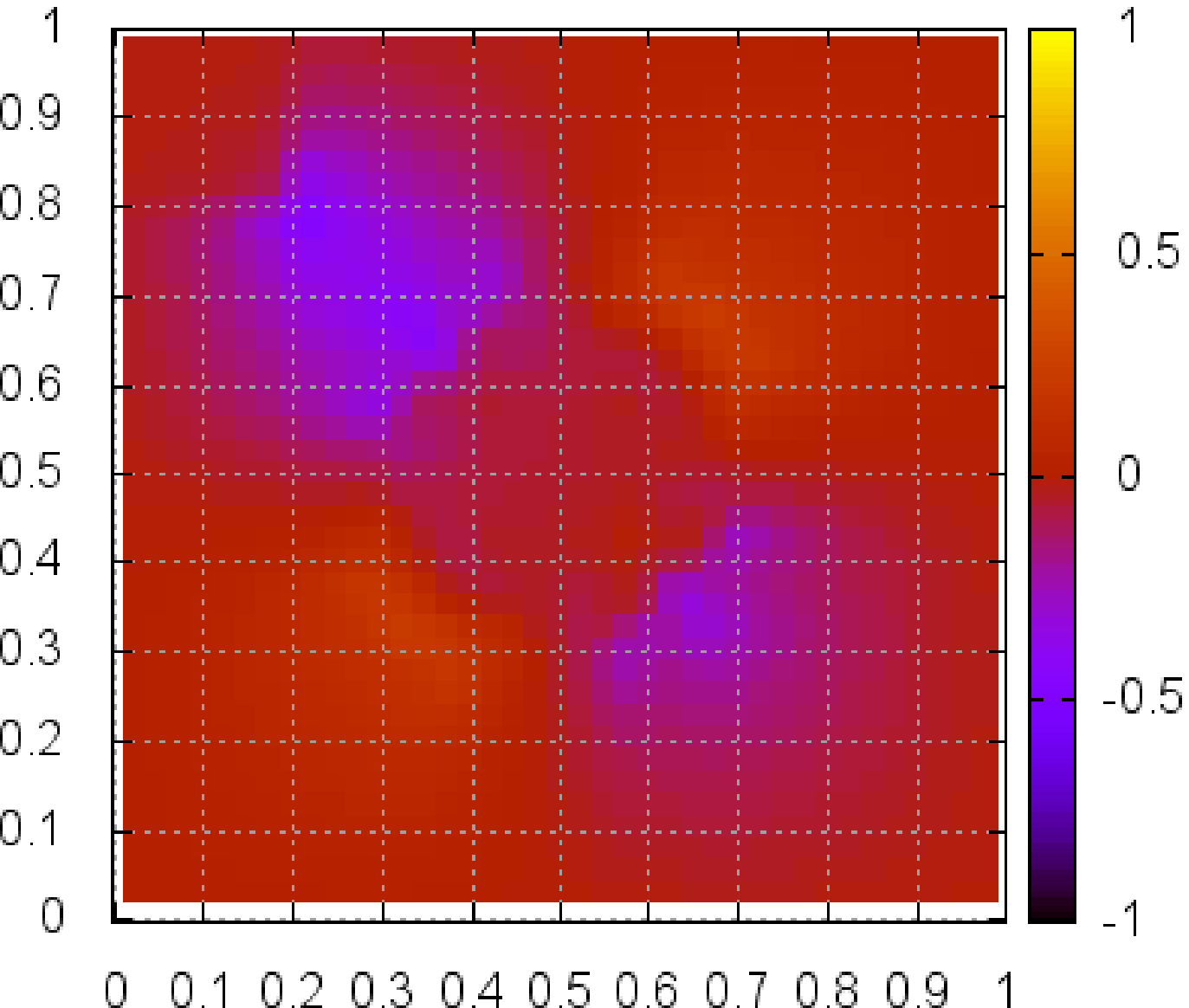}}
\hfill
\subfigure[no noise - posterior mean]{
\includegraphics[width=0.3\textwidth]{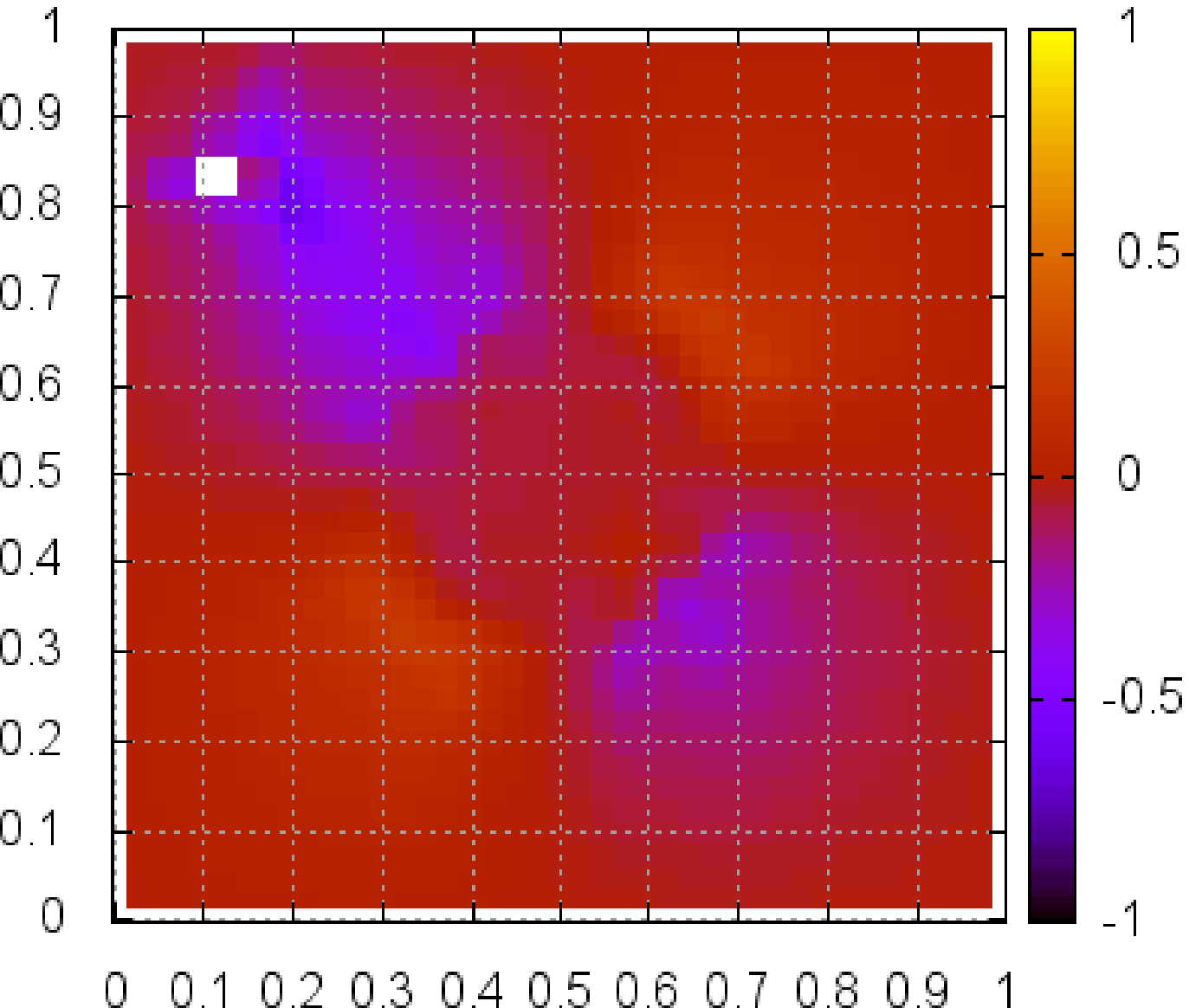}}
\hfill
\subfigure[SNR=40 - posterior mean]{
\includegraphics[width=0.3\textwidth]{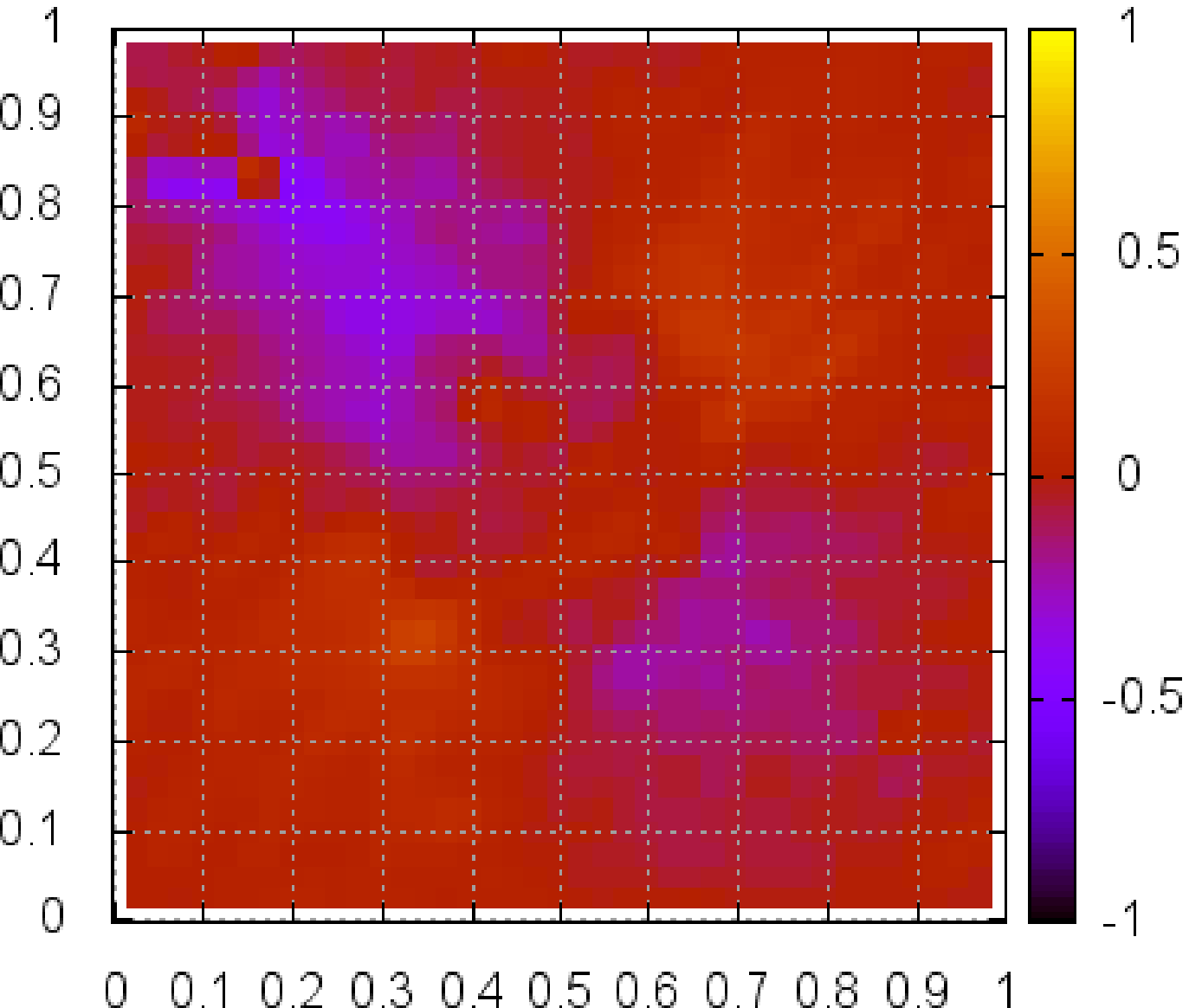}}
\caption{Example 2: Comparison of shear stress' $\sigma_{xy}$  spatial distribution with the posterior means obtained when data have no noise and for SNR=$40dB$.}
\label{fig:one_ellipse_shear} 
\end{figure}

Figure \ref{fig:one_ellipse_sl2} depicts the learned values of the the parameters $\bl=\{\lambda_e^2\}$ (\refeq{eq:pece}) which express the magnitude of model error over each element of the domain. Both in the absence of noise and when SNR=$40dB$, the algorithm clearly identifies a significant model error in the region on the top-left corner. It is noted that that the $\lambda_e^2$ values in this region are $2$ to $4$ orders of magnitude larger than in the rest of the problem domain.
Despite the model inadequacy the algorithm correctly identifies the presence of the circular inclusion as it can be seen in Figure \ref{fig:one_ellipse} and more clearly in Figure \ref{fig:one_ellipse_diag} which shows  the elastic modulus variation along the diagonal from $(0,1)$ to $(1,0)$.
It is particularly interesting to note that even though the isotropic elastic constitutive model endowed in the inversion scheme is inadequate at least for a subdomain of the problem, the proposed scheme can correctly identify the stresses (pressure and shear)  in the whole domain 
 as it can be seen in Figures \ref{fig:one_ellipse_pressure} and \ref{fig:one_ellipse_shear}. These depict the ground truth in comparison with the posterior means obtained with no noise and for SNR=$40 dB$. The posterior quantiles (which are omitted herein for economy of space) fully envelop the ground truth.

\begin{figure}[!h]
\psfrag{distance}{distance along diagonal}
\psfrag{modulus}{Elastic modulus $E$}
\subfigure[no noise]{
\includegraphics[width=0.48\textwidth]{FIGURES/one_ellipses_diag1_snr=0.eps}
\label{diagc}}
\hfill
\subfigure[SNR=40]{
\includegraphics[width=0.48\textwidth]{FIGURES/one_ellipses_diag1_snr=40.eps}
\label{diagd}
}
\caption{Example 2: Posterior statistics of the elastic modulus distribution along the diagonal from $(0,1)$ to $(1,0)$ }
 \label{fig:one_ellipse_diag}
\end{figure}

\section{Conclusions}
While existing stochastic (Bayesian) strategies for the solution of inverse problems associated with the identification of material properties in biomechanics are able to account for various sources of uncertainty in the problem, they are generally deficient in terms of assessing model fidelity. We proposed an intrusive formulation that incorporates  the various model equations in the likelihood  (posterior)   and is capable of inferring model discrepancies from noisy displacement data. In contrast to direct methods, it does not require imputations of strains nor their derivatives.
It provides probabilistic confidence metrics (credible intervals) that can be very useful to the analyst as well as probabilistic estimates of the (unobserved) stresses/pressures. We discussed a scalable computational framework which can be greatly accelerated by employing a multi-resolution strategy. The latter could be utilized in order to propose adaptively, refinements of the discretized domain which we intend to explore in the future. Current investigations also involve extending this approach to dynamic settings  where the parameter vector should include velocities and accelerations in addition to displacements, and the model equations should  include the time-integration scheme adopted.

\clearpage
\newpage
\label{sec:appendix}
{\bf Appendix: Maximization with respect to $\bl$}

This section describes the computations involved during the Maximization step of the EM algorithm described in section \ref{sec:meth}.
In particular according to Equations (\ref{eq:q}), (\ref{eq:estep2}), (\ref{eq:ss}) and the prior model in \refeq{eq:priorl}, this entails a maximization with respect to $\bl=\{\lambda_e^2\}_e$ of:
\bee
\begin{array}{ll}
 Q(\mathbf{\Lambda}^{(j)}, \mathbf{\Lambda}) &=-\frac{n_{\sigma}}{2}\sum_{e=1}^{n_{el}}\log \lambda_e^2-\frac{1}{\lambda_e^2} \sum_{e=1}^{n_{el}}\Phi_e^{(j)}+\log p(\bl) \\
& =-\frac{n_{\sigma}}{2}\sum_{e=1}^{n_{el}}\log \lambda_e^2-\frac{1}{\lambda_e^2} \sum_{e=1}^{n_{el}}\Phi_e^{(j)} -\frac{1}{2}\mathbf{Z}^T \mathbf{W} \mathbf{Z}
\end{array}
\eee
It is reminded that the vector $\mathbf{Z}=\{z_e\}_{e=1}^{n_{el}}$ contains the log values of $\bl$ i.e. $z_e =\log \lambda_e^2$.
Rather than solving an optimization in the $n_{el}$-dimensional space at each iteration $j$, we perform successive updates of each $\lambda_e^2$ or $z_e$ while keeping the remaining fixed. This {\em incremental} version of the EM algorithm entails performing $n_{el}$ optimizations of one-dimensional functions. We propose carrying out the latter task with respect to $z_e $ (as they are allowed to take any value on the real axis in contrast to $\lambda_e^2$ which must be positive) and employ  a  standard Newton-Raphson scheme. This requires the first and second order derivatives of the objective function above which are given by:
\bee
\frac{\pa Q(\mathbf{\Lambda}^{(j)}, \mathbf{\Lambda})}{\pa z_e} = -\frac{3}{2}+\frac{\Phi_e^{(j)}}{2} e^{-z_e} -\frac{z_e-\mu_{z_e}}{\sigma_{z_e}^2}
\eee
and:
\bee
\frac{\pa^2 Q(\mathbf{\Lambda}^{(j)}, \mathbf{\Lambda})}{\pa z_e^2} = -\frac{\Phi_e^{(j)}}{2} e^{-z_e} -\frac{1}{\sigma_{z_e}^2}
\eee
where:
\bee
\begin{array}{l}
\sigma_{z_e}^2=1/W_{e,e} \\
\mu_{z_e}= -\frac{1}{W_{e,e}} \sum_{k\ne e} W_{e,k} z_k
\end{array}
\eee
It can be easily seen that the second derivative is always, strictly negative $\frac{\pa^2 Q(\mathbf{\Lambda}^{(j)}, \mathbf{\Lambda})}{\pa z_e^2} <0$  and therefore the problem is convex.

\section*{References}
\bibliographystyle{jphysicsB}
\bibliography{paper}

\end{document}